\documentclass[aps,pra,reprint,superscriptaddress,nofootinbib,floatfix]{revtex4-2}
\usepackage{amsmath,amssymb,bm}
\usepackage{graphicx}
\usepackage{tikz}
\usetikzlibrary{arrows.meta,calc}

\newcommand{\sgn}{\operatorname{sgn}}
\newcommand{\diag}{\operatorname{diag}}
\newcommand{\tr}{\operatorname{tr}}
\newcommand{\Perm}{\mathcal{P}}

\begin{document}

\title{Solitons as conversion events in quadratic crystals}

\author{Francesco Beccarini}
\author{Claudio Conti}
\affiliation{Dipartimento di Fisica, Sapienza Universit\`a di Roma, Piazzale Aldo Moro 5, 00185 Rome, Italy}

\date{\today}

\begin{abstract}
In a $\chi^{(2)}$ crystal with temporal walk-off and negligible dispersion, signal, idler and pump pulses obey the integrable three-wave interaction. For a pump of intermediate group velocity we describe each soliton as a conversion event: the fusion of a signal--idler pair into a pump pulse, or the decay of a pump pulse, at a definite depth in the crystal. The asymptotic $\mathrm{sech}$ pulses, usually treated with three separate $2\times2$ scattering problems, are the legs of the event, a single pole of the $3\times3$ problem. For one event we give the conversion curve in closed form and the minimum crystal length; for several events, the rule by which they displace each other. A pump pulse is unstable, and its growing mode is the tail of a decay event: a seed carrying a fraction $\varepsilon$ of the photons along this mode makes the pulse decay after $\Gamma^{-1}\ln(c_+/\varepsilon)$, with $\Gamma$ the conversion rate and $c_+$ of order one. Unmatched signal and idler pulses form the resonance of Kaup, Reiman and Bers; we show that it is a fusion followed by a decay, each within a conversion length, and give its pump photon number in closed form. Since the delay between the input pulses sets the depth of the vertex, these quantities can be read on the output face of a single crystal; we give the scales for periodically poled KTiOPO$_4$.
\end{abstract}

\maketitle

\section{Introduction}\label{sec:intro}

Three-wave mixing in quadratic media underlies sum and difference frequency conversion, second-harmonic generation, and parametric amplification \cite{Boyd}; it is the standard tool for broadband frequency conversion \cite{Cerullo03}. For plane waves, the energy flows periodically between harmonic waves \cite{Armstrong62}. For pulses the waves walk off each other, the energy exchange is not periodic, and the emerging states depend on the shape, duration, fluency, and delay of input pulses.

Without dispersion, diffraction, and cubic nonlinearity, the electric field envelopes obey the three-wave interaction system (TWI), which is integrable by the inverse scattering transform \cite{ZM73,ZM75,Kaup76}. Kaup, Reiman and Bers described TWI evolution for localized data \cite{KRB79}: a pulse above the so-called area threshold decays into the other two waves, after a triggering collision. In optics, the compression of the sum-frequency pulse generated by pre-delayed fundamental pulses was found numerically and observed \cite{WangDragila90,Stabinis91,Umbrasas95,Dubietis98}, and explained by soliton solutions of the TWI \cite{IbragimovStruthers96,IbragimovStruthers97}. A similar approach showed that the output of a parametric amplifier is fixed by the soliton content of the pump \cite{IbragimovKaupStruthers98,IbragimovKaup97,Ibragimov99}. Baronio and co-workers observed the spatial counterpart with non-collinear beams in KTiOPO$_4$, where the transverse coordinate replaces time: above an intensity threshold the generated sum-frequency beam decays at a definite depth and the fundamental beams leave displaced \cite{Baronio09}. In the same geometry they observed an ensemble of non-interacting bright--dark--bright triads predicted from the spectral data of the input beams \cite{Baronio11}. They noted that the sum-frequency pulse generated by two colliding pulses decays back after a short distance \cite{Baronio10}, and reviewed the instability of the solitons of the intermediate-velocity wave together with spectral and numerical methods for the direct problem \cite{Lombardo11}.

In this literature the solitons are the asymptotic profiles, TWI solitons \cite{IbragimovStruthers97,Ibragimov99}. For separated pulses the $3\times3$ scattering problem factorizes into three $2\times2$ Zakharov--Shabat problems \cite{KRB79,Ibragimov99}, and a TWI soliton is a $\mathrm{sech}$ pulse of one wave, usually described as supported by the group-velocity mismatch \cite{Ibragimov99}, as distinct from the quadratic solitons sustained by diffraction or dispersion \cite{Buryak02}. The description is exact asymptotically but hides two facts. Away from other waves each envelope obeys a linear transport equation: the shape of an asymptotic pulse is sustained by no balance, and is the imprint of an interaction that took place elsewhere. The signal, idler and pump pulses of one interaction are a single eigenvalue of the $3\times3$ problem, which the factorization splits into eigenvalues of different waves. The interaction takes place where the pulses overlap, where the factorization fails \cite{Ibragimov99}.

In this reasoning, the TWI soliton is the conversion event together with its ``legs'': a fusion or a decay at a point of the crystal, i.e. the vertex, and the asymptotic pulses that leave it. The two are parts of one single solution: the vertex fixes the shape, timing and phases of the legs, and the legs carry out of the crystal what can be measured of the vertex, the profiles of the soliton.

Several aspects in this picture are known from the cited literature: the classification into fusion and decay, the area threshold, the instability of a pulse of the intermediate-velocity wave, the pairing of signal and idler eigenvalues required for complete up-conversion, the transient up-conversion of unpaired pulses, with a time delay logarithmic in the mismatch, and the decay of a pump pulse triggered by noise, with a lifetime logarithmic in the noise energy \cite{KRB79,Lombardo11}; the compression ratio, the optimal intensity ratio and the time shifts of the fundamental pulses in sum-frequency generation \cite{IbragimovStruthers96,IbragimovStruthers97}; the decay back of the sum-frequency pulse and its observation with beams \cite{Baronio09,Baronio10,Lombardo11}; and the trapping of signal and idler by the gain of a pump pulse, with closed-form growing modes for an undepleted pump \cite{Tsang06,Gorbach23,Hamrouni24}.

In this manuscript, we revisit the theory of the three-wave interaction and derive the following results. 
We show that the growing mode of a pump pulse is the tail of a decay event, so that the depth of the decay is fixed by the weight of the seed along the mode. This explains the transient up-conversion of unpaired pulses: a signal and an idler pulse whose poles are split by $|\theta|$ fuse into a pump pulse seeded by $c_+|\theta|^2$, and the resonance they form is exactly a fusion followed by a decay: the pump is formed within a conversion length at the collision, survives unchanged over $\tau/2$, with $\tau$ the delay of Ref.~\cite{KRB79}, and decays within a conversion length, with a pump photon number in closed form for any splitting. In Ref.~\cite{KRB79} the pump was described as growing and decaying over the whole delay (Sec.~\ref{sec:resonance}). Back-conversion of a matched fusion is the same decay, seeded by whatever splits the poles. We also derive the exact conversion curve of an event and the minimum crystal length; the rule by which the $N$ events of an $N$-soliton displace each other along the characteristics of definite waves; and, for separated input pulses, the pairing condition in closed form in their durations and energies. These quantities concern the region where the waves overlap and are not contained in the three $2\times2$ problems of the asymptotic description. As detailed below, these results can be tested experimentally by varying the delay between the input pulses (Sec.~\ref{sec:scales}).

The outline of the manuscript is the following: Section~\ref{sec:optics} derives the model in its propagation form, with the depth in the crystal as evolution parameter and the field on the input face as initial datum. Section~\ref{sec:model} collects the Lax pair description of the model and the conserved charges, derived in Appendix~\ref{app:charges}. Section~\ref{sec:solitons} contains the results. For a single event the conversion curve fixes the depth of the conversion and the minimum crystal length beyond which the faces carry the asymptotic pulses (Secs.~\ref{sec:onesol}--\ref{sec:faces}); several events displace each other along the characteristics of definite waves (Secs.~\ref{sec:two} and \ref{sec:shifts}). Secs.~\ref{sec:pure} and \ref{sec:resonance} treat the pump pulse and the resonance: a pump pulse is unstable, its growing mode is the tail of a decay event, and an unmatched signal--idler pair forms a resonance, a fusion followed by a decay, with a lifetime logarithmic in the splitting of their poles. Section~\ref{sec:scales} gives the scales for type-II interaction in periodically poled KTiOPO$_4$ at extended phase matching \cite{KoenigWong04} and the measurements that access these quantities. Conclusions are drawn in Sec.~\ref{sec:conclusions}.

\section{From the optical system to the model}\label{sec:optics}

\subsection{Coupled-envelope equations}

We consider a $\chi^{(2)}$ medium in which three carriers with fixed polarizations propagate collinearly along $z$, either as plane waves or in a single transverse mode of a channel waveguide: the signal at $\omega_1$, the idler at $\omega_2$ and the pump at $\omega_3=\omega_1+\omega_2$, with $k_j=n_j\omega_j/c$ and phase mismatch $\Delta k=k_3-k_1-k_2$. With the field $E=\tfrac12\sum_jA_j(z,t)\,e^{i(k_jz-\omega_jt)}+{\rm c.c.}$ and the nonlinear polarization $P_{\rm NL}=2\varepsilon_0d_{\rm eff}E^2$, in the slowly varying envelope approximation and to first order in the bandwidth, $k(\omega_j+\Omega)=k_j+\beta_{1,j}\Omega+\tfrac12\beta_{2,j}\Omega^2+\dots$ with $\beta_{1,j}=1/v_{g,j}$ the inverse group velocity, the phase-matched components give \cite{Boyd}
\begin{equation}\label{eq:envelopes}
\begin{aligned}
(\partial_z+\beta_{1,1}\partial_t)A_1&=i\alpha_1A_3\bar A_2e^{i\Delta kz},\\
(\partial_z+\beta_{1,2}\partial_t)A_2&=i\alpha_2A_3\bar A_1e^{i\Delta kz},\\
(\partial_z+\beta_{1,3}\partial_t)A_3&=i\alpha_3A_1A_2e^{-i\Delta kz},
\end{aligned}
\end{equation}
with $\alpha_j=\omega_jd_{\rm eff}/(n_jc)$: one pump photon is exchanged with one signal--idler pair.

\subsection{Photon-flux variables and reduction}

With
\begin{equation}
a_j=\sqrt{\frac{n_j\varepsilon_0c}{2\hbar\omega_j}}\,A_j,\qquad |a_j|^2=\frac{I_j}{\hbar\omega_j}\quad(\text{photons s}^{-1}\text{m}^{-2}),
\end{equation}
the three couplings become equal,
\begin{gather}
(\partial_z+\beta_{1,1}\partial_t)a_1=i\gamma\,a_3\bar a_2e^{i\Delta kz},\ \ \text{etc.},\\
\gamma=d_{\rm eff}\sqrt{\frac{2\hbar\,\omega_1\omega_2\omega_3}{\varepsilon_0c^3n_1n_2n_3}}\label{eq:gamma}
\end{gather}
(divided by $\sqrt{A_{\rm eff}}$, with the mode-overlap integral, in a waveguide). The constant $\hbar$ enters only through the photon normalization and cancels from every intensity computed below. The symmetric coupling reflects the Hamiltonian structure, and the Manley--Rowe relations are local conservation laws,
\begin{equation}
\partial_z\big(|a_1|^2+|a_3|^2\big)+\partial_t\big(\beta_{1,1}|a_1|^2+\beta_{1,3}|a_3|^2\big)=0,
\end{equation}
and the same with $1\to2$: every pump photon destroyed creates one signal and one idler photon.

\emph{Comoving frame.} With the retarded time $\tau=t-\beta_{\rm ref}z$, $\partial_z+\beta_{1,j}\partial_t\to\partial_z+\delta_j\partial_\tau$, $\delta_j=\beta_{1,j}-\beta_{\rm ref}$ (group-velocity mismatch, temporal walk-off). The model depends on the $\delta_j$ only through their differences, so $\beta_{\rm ref}$ is arbitrary; below we often take the pump frame, $\beta_{\rm ref}=\beta_{1,3}$.

\emph{Phase mismatch.} Setting $a_j=b_je^{i(p_j\tau+q_jz)}$ with $p_3=p_1+p_2$, $q_3=q_1+q_2-\Delta k$ and $q_j=-\delta_jp_j$, the $b_j$ obey the phase-matched equations provided
\begin{equation}
p_1(\delta_3-\delta_1)+p_2(\delta_3-\delta_2)=\Delta k ,
\end{equation}
which is always solvable for distinct group velocities: in the absence of group-velocity dispersion a constant $\Delta k$ amounts to a shift of the carrier frequencies (keeping $\omega_3=\omega_1+\omega_2$) to exact phase matching, and does not change the model.

\emph{Sign.} Finally, $u_1=b_1$, $u_2=b_2$, $u_3=-b_3$ turns $+i\gamma$ into $-ig$, $g=\gamma$, in all three equations. The sign is conventional: it gives the form \eqref{eq:3wi}, in which the Lax pair of Sec.~\ref{sec:lax} is written.

\subsection{Dictionary}

Table~\ref{tab:dictionary} reports the meaning of the variables of Eq.~\eqref{eq:3wi}, the form adopted below. We shall mostly use the \emph{propagation reading}, in which the evolution parameter of the model is the depth $z$ in the crystal and the spatial variable is the retarded time $\tau$; the initial datum is then the field on the input face, $z=0$, and the crystal length is the final value of the evolution parameter. Multiplying by $v_{g,j}$ and using photon line densities $c_j=a_j/\sqrt{v_{g,j}}$ one obtains the same form in the \emph{time reading}, with $t$ the laboratory time, $x=z$, $v_j=v_{g,j}$ and $g=\gamma\sqrt{v_{g,1}v_{g,2}v_{g,3}}$; a finite crystal then requires genuine boundary conditions at the input and output faces, which is one reason why the propagation reading is the natural one.

\begin{table}[t]
\caption{\label{tab:dictionary}Correspondence between the model, Eq.~\eqref{eq:3wi}, and the optical system in the propagation reading.}
\begin{ruledtabular}
\begin{tabular}{ll}
model & optics (propagation reading)\\
\hline
$t$ & depth in the crystal $z$\\
$x$ & retarded time $\tau$\\
$v_j$ & $\delta_j=1/v_{g,j}-1/v_{g,{\rm ref}}$ (s/m)\\
$w_j$ & group-velocity mismatches (s/m)\\
$u_j$ & photon-flux envelopes (phases as above)\\
$g$ & $\gamma$, Eq.~\eqref{eq:gamma}\\
$N_j$ & photons crossing a transverse plane (per unit area)\\
$H$ & generator of $z$-translations\\
$P$ & generator of $\tau$-translations (frequency detuning)\\
\end{tabular}
\end{ruledtabular}
\end{table}

\emph{Hamiltonian structure.} In the time reading $H$ is the field energy in the frame rotating at the carriers, i.e., measured from $\sum_j\hbar\omega_jN_j=\hbar(\omega_1\mathcal Q_1+\omega_2\mathcal Q_2)$, which is conserved, and $P$ is the momentum measured from the carriers. In the propagation reading the roles are exchanged: $H$ generates $z$-translations, i.e., it is the momentum flux through the plane $z$ measured from the carriers in the comoving frame, and $P$ is the frequency-detuning flux, since with $a(\tau)=\int\frac{d\Omega}{2\pi}\tilde a(\Omega)e^{-i\Omega\tau}$ one has $-i\partial_\tau\to-\Omega$, a component at $\omega_j+\Omega$ carrying the extra wave vector $\delta_j\Omega$ in the comoving frame.

\emph{Transverse reading.} The same model also describes continuous-wave beams in a planar waveguide, with $x$ the transverse coordinate, $t=z$ and $v_j=\tan\rho_j$ the tangents of the energy-flow directions (spatial walk-off). All the results below apply to that setting with $\tau\to x$; we keep the temporal language throughout.

\section{Integrability}\label{sec:model}

\subsection{Model}

The fields are the signal $u_1$, the idler $u_2$ and the pump $u_3$; the parameters are the velocities $v_j$ and the coupling $g>0$:
\begin{equation}\label{eq:3wi}
\begin{aligned}
(\partial_t+v_1\partial_x)u_1&=-ig\,\bar u_2u_3,\\
(\partial_t+v_2\partial_x)u_2&=-ig\,\bar u_1u_3,\\
(\partial_t+v_3\partial_x)u_3&=-ig\,u_1u_2,
\end{aligned}
\end{equation}
with fundamental Poisson brackets
\begin{equation}
\{\bar u_j(x), u_k(y)\}=i\delta_{jk}\delta(x-y).
\end{equation}
We use the relative velocities and the constants
\begin{gather}
w_1=v_2-v_3,\quad w_2=v_3-v_1,\quad w_3=v_1-v_2,\\
\kappa=-\frac{g^2}{w_1w_2w_3},\qquad \zeta_j(x,t)=v_jx+v_kv_l\,t,
\end{gather}
$(j,k,l)$ being a permutation of $(1,2,3)$.

\emph{Boundary conditions.} The three fields decrease rapidly for $|x|\to\infty$ at fixed $t$ [$u_j(\cdot,t)\in L^1(\mathbb R)$ suffices for the existence of the transition matrix of Appendix~\ref{app:charges}]. The class is preserved by the evolution: Eq.~\eqref{eq:3wi} is hyperbolic with characteristic speeds $v_j$, so a field vanishing outside $[a,b]$ at $t=0$ vanishes outside $[a+t\min_jv_j,\ b+t\max_jv_j]$. In the propagation reading these are finite-energy pulses on the input face. Non-vanishing backgrounds are excluded: a constant undepleted pump, $u_3\to{\rm const}\neq0$, $u_{1,2}\to0$ (continuous-wave-pumped amplifier), is a stationary solution but parametrically unstable, and requires the inverse scattering transform with nonzero boundary conditions, which is not treated here. For the same reason the parametric solitons with a nonvanishing component and locked velocity studied in Refs.~\cite{Degasperis06,Conforti06} are outside the class considered below.

\subsection{Lax pair}\label{sec:lax}

Equation~\eqref{eq:3wi} is the compatibility condition of the linear problem $\psi_x=U\psi$, $\psi_t=V\psi$, $\psi\in\mathbb{C}^3$ \cite{ZM73,ZM75,Kaup76,KRB79}, with
\begin{gather}
U=-i\lambda A+Q,\qquad V=-i\lambda B+\tilde Q,\\
A=\diag(v_1,v_2,v_3),\qquad B=\diag(v_2v_3,\,v_3v_1,\,v_1v_2),\\
\tilde Q_{jk}=-v_l\,Q_{jk}\quad (l\neq j,k),
\end{gather}
\begin{equation}
Q=\begin{pmatrix}
0 & \rho_3\bar\beta_3\,\bar u_3 & \rho_2\beta_2\,\bar u_2\\
\beta_3\,u_3 & 0 & \beta_1\,u_1\\
\beta_2\,u_2 & \rho_1\beta_1\,\bar u_1 & 0
\end{pmatrix},
\end{equation}
\begin{gather}
\beta_1=\frac{g}{\sqrt{|w_2w_3|}},\quad
\beta_2=\frac{g}{\sqrt{|w_3w_1|}},\quad
\beta_3=-\frac{i g\,\sgn w_3}{\sqrt{|w_1w_2|}},\\
\rho_1=\sgn(w_2w_3),\quad \rho_2=\sgn(w_3w_1),\quad \rho_3=-\sgn(w_1w_2).
\end{gather}
Then
\begin{equation}\label{eq:zc}
U_t-V_x+[U,V]=0\ \ \forall\lambda\quad\Longleftrightarrow\quad\text{Eq.~\eqref{eq:3wi}}.
\end{equation}
The potential obeys the reality condition
\begin{gather}
Q^\dagger=-SQS,\label{eq:reality}\qquad U(\bar\lambda)^\dagger=-S\,U(\lambda)\,S,\\
S=\diag(\sgn w_1,\ \sgn w_2,\ -\sgn w_3).
\end{gather}
The reality condition is a symmetry reduction of the algebra in which the Lax pair takes values. A priori $U$ and $V$ are $\mathfrak{gl}(3)$-valued; for real $\lambda$, Eq.~\eqref{eq:reality} reads $U^\dagger S+SU=0$, i.e., $U$ preserves the Hermitian form $\psi^\dagger S\psi$. Its trace, $-i\lambda(\tr A/3)\,\mathbb I$, drops out of Eq.~\eqref{eq:zc}, and the traceless part lies in $\mathfrak{su}(3)$ for $S=\pm\mathbb I$, i.e., when $v_3$ lies between $v_1$ and $v_2$ (pump of intermediate group velocity). A sign pattern that is not homogeneous gives $\mathfrak{su}(2,1)$, which corresponds to the other orderings of the group velocities and is not considered here. The reduction is a property of the Lax pair, not a symmetry of the model: the conserved charges of Sec.~\ref{sec:charges} generate the phase rotations of the fields and the translations, and the Poisson structure is that of $\mathfrak{gl}(3)$ (Appendix~\ref{app:charges}).

\subsection{Conserved charges}\label{sec:charges}

The Poisson brackets of $U$ are ultralocal and given by Yang's rational $\mathfrak{gl}(3)$ $\mathfrak r$-matrix, $\mathfrak r(\lambda)=\kappa\Perm/\lambda$, so that the traces of the powers of the monodromy matrix are in involution (Appendix~\ref{app:charges}). The simplest local conserved charges are
\begin{align}
    &\mathcal Q_1=N_1+N_3= \int dx \left[|u_1|^2+|u_3|^2 \right], \\
    &\mathcal Q_2=N_2+N_3=  \int dx \left[|u_2|^2+|u_3|^2\right],\\
    &H=\int dx\left[-i\sum_j v_j\bar u_j\partial_xu_j+g\left(\bar u_1\bar u_2u_3+u_1u_2\bar u_3\right)\right], \\
    &P=-i\sum_j\int dx\, \bar u_j\partial_xu_j .
\end{align}
Because the problem is $3\times3$ and $\det\mathcal T=1$, each order of the expansion of the monodromy at $\lambda=\infty$ contains two independent charges: the two Manley--Rowe charges $\mathcal Q_{1,2}$ at first order, and $P$ and $H$ at second order. The Manley--Rowe charges are the ones used below to count photons on the boundaries.

\section{Solitons as conversion events}\label{sec:solitons}

Hereafter the pump has the intermediate group velocity, $S=\mathbb I$ and $v_1<v_3<v_2$, so that $w_1,w_2>0$ and $w_3=-W$ with
\begin{equation}
W=w_1+w_2=|v_2-v_1|,\qquad r=\frac{w_2}{W}\in(0,1).
\end{equation}
In the optical variables $W=|\beta_{1,1}-\beta_{1,2}|$ is the signal--idler group-velocity mismatch; $r=1/2$ when the pump group velocity is the mean of signal and idler velocities. We work in the pump frame, $v=(-w_2,w_1,0)$, unless stated otherwise.

\subsection{Dressing and the $N$-soliton solution}\label{sec:dressing}

The solitons are obtained by dressing the zero solution, $\Psi_0=e^{-i\lambda Z}$, $Z=Ax+Bt=\diag(\zeta_1,\zeta_2,\zeta_3)$ \cite{ZS79}; Darboux-dressing solutions of the three-wave equations are given in Ref.~\cite{DegasperisLombardo06}. Given $N$ poles $\mu_k=\xi_k+i\eta_k\in\mathbb{C}\setminus\mathbb{R}$ and polarization vectors $\bm n_k^0\in\mathbb{C}^3$, let
\begin{gather}
\bm n_k=e^{-i\mu_kZ}\bm n^0_k,\qquad M_{kl}=\frac{\bm n_k^\dagger S\,\bm n_l}{\mu_l-\bar\mu_k},\label{eq:gram}\\
\chi(\lambda)=\mathbb{I}-\sum_{k,l=1}^N\frac{\bm n_k\,(M^{-1})_{kl}\,\bm n_l^\dagger S}{\lambda-\bar\mu_l},\qquad\Psi=\chi\Psi_0 .
\end{gather}
Then $Q=-i\big[A,\sum_{k,l}\bm n_k(M^{-1})_{kl}\bm n_l^\dagger S\big]$, and with $u_1=Q_{23}/\beta_1$, $u_2=Q_{31}/\beta_2$, $u_3=Q_{21}/\beta_3$ the fields are ratios of determinants,
\begin{gather}
u_1=\frac{iw_1}{\beta_1}\frac{G_{23}}{G},\qquad u_2=\frac{iw_2}{\beta_2}\frac{G_{31}}{G},\nonumber\\
u_3=-\frac{iw_3}{\beta_3}\frac{G_{21}}{G},\label{eq:Nsol}
\end{gather}
where $G=\det M$ and $G_{jk}=\det\begin{pmatrix}M&\bar{\bm n}^{(k)}s_k\\ \bm n^{(j)T}&0\end{pmatrix}$, $\bm n^{(j)}=(n_{1,j},\dots,n_{N,j})^T$ collecting the $j$th components of the $N$ polarization vectors, and $s_j$ the diagonal entries of $S$.

\emph{Physical parameters.} Since $\zeta_j$ is linear in $(x,t)$, every polarization vector can be written as
\begin{equation}\label{eq:physpar}
n_{k,j}(x,t)=e^{i\phi_{k,j}}\exp\!\big[-i\mu_k\,\zeta_j(x-x_k,\,t-t_k)\big],
\end{equation}
so that each soliton is labeled by six real numbers with a direct meaning. The point $(x_k,t_k)$ is the \emph{nominal vertex} at which the three components of $\bm n_k$ have equal modulus; $\eta_k$ fixes the pulse durations and amplitudes; $\xi_k$ is a common frequency detuning, since by Eq.~\eqref{eq:physpar} the signal, idler and pump fields of soliton $k$ carry the phases $e^{-i\Omega_j\tau}$ with
\begin{equation}
\Omega_1=\xi_kw_1,\qquad \Omega_2=\xi_kw_2,\qquad \Omega_3=-\xi_kw_3=\Omega_1+\Omega_2,
\end{equation}
consistent with energy conservation; and two relative phases $\phi_{k,j}$ fix the phases of the fields. The coupling $g$ enters only through the prefactors $\beta_j$ in Eq.~\eqref{eq:Nsol}: it sets the amplitudes, not the geometry of the solution.

\subsection{One soliton: fusion and decay}\label{sec:onesol}

\begin{figure*}[t]
\includegraphics[width=\textwidth]{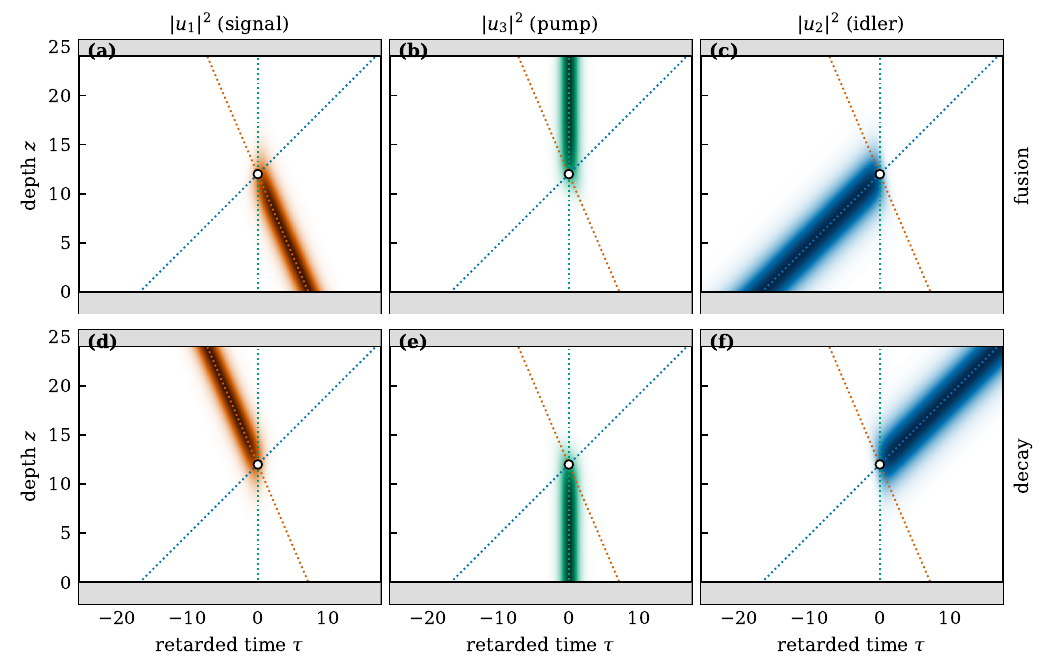}
\caption{\label{fig:events}Single fusion (top row, $\eta=0.5$) and decay (bottom row, $\eta=-0.5$) events in a crystal of length $L=24$ (input and output faces: black lines), for $r=0.3$ ($v=(-0.6,1.4,0)$, $g=1$). Photon fluxes $|u_j|^2$ of signal, pump and idler in the $(\tau,z)$ plane; dotted lines: characteristics through the vertex (open circle), placed at $z_0=12$, $\tau_0=0$.}
\end{figure*}

For $N=1$, $\mu=\xi+i\eta$ and $D=\sum_js_j|n_j|^2$,
\begin{equation}\label{eq:1sol}
\begin{aligned}
u_1&=\frac{2\eta}{g}\,w_1\sqrt{|w_2w_3|}\;\frac{s_3\,n_2\bar n_3}{D},\\
u_2&=\frac{2\eta}{g}\,w_2\sqrt{|w_3w_1|}\;\frac{s_1\,n_3\bar n_1}{D},\\
u_3&=-\frac{2i\eta}{g}\,|w_3|\sqrt{|w_1w_2|}\;\frac{s_1\,n_2\bar n_1}{D}.
\end{aligned}
\end{equation}
Away from the vertex, on the ray $x\simeq v_it$ ($(i,j,k)$ cyclic),
\begin{gather}
|u_i|\to\frac{|\eta w_i|\sqrt{|w_jw_k|}}{g}\,\mathrm{sech}\big[\eta w_i(x-v_it)+\ln|n^0_j/n^0_k|\big],\label{eq:leg}\\
\int\!|u_i|\,dx=\frac{\pi}{g}\sqrt{|w_jw_k|},\qquad \mathcal N\equiv N_i=\frac{2|\eta\,w_1w_2w_3|}{g^2}.\label{eq:Nevent}
\end{gather}
The solution is a Y-shaped junction in the $(\tau,z)$ plane (Fig.~\ref{fig:events}): the pump leg on one side of the vertex, the signal and idler legs on the other. For $S=\mathbb I$,
\begin{gather}
\eta>0:\ u_1+u_2\to u_3\quad(\text{fusion}),\nonumber\\
\eta<0:\ u_3\to u_1+u_2\quad(\text{decay}),
\end{gather}
\cite{ZM75,Kaup76,KRB79}. The sign of $\eta$ is not a convention. The dressing factor of one pole, $\chi=\mathbb I-\frac{\mu-\bar\mu}{\lambda-\bar\mu}\Pi$ with $\Pi=\bm n\bm n^\dagger S/(\bm n^\dagger S\bm n)$, is proportional to $\Pi+\frac{\lambda-\bar\mu}{\lambda-\mu}(\mathbb I-\Pi)$: the same solution has a rank-one zero at $\mu$ and a rank-two zero at $\bar\mu$, and $\eta$ fixes in which half-plane the rank-one zero lies. Each leg is a $\mathrm{sech}$ pulse of duration $T_i=1/|\eta w_i|$ and of area $\pi$ in the units of its own $2\times2$ Zakharov--Shabat block, $\beta_i\int|u_i|dx=\pi$, and all three legs carry the same photon number $\mathcal N$: the soliton converts $\mathcal N$ signal--idler pairs into $\mathcal N$ pump photons, or vice versa, \emph{completely}. Since $T_3/T_{1,2}=w_{1,2}/W<1$, the pump leg is always the shortest and the most intense, $|u_3|^2_{\max}/|u_{1,2}|^2_{\max}=W/w_{1,2}$; the fusion soliton is the soliton mechanism of pulse compression in sum-frequency generation \cite{KRB79,IbragimovStruthers96,IbragimovStruthers97}. The pump phase is locked to the phases of signal and idler, $\arg u_3=\arg(u_1u_2)+{\rm const}$.

\emph{Conversion curve.} The photon numbers along the crystal follow a universal law. In the pump frame, with $y=|n_2/n_1|^2$ as integration variable, the integral of $|u_3|^2$ from Eq.~\eqref{eq:1sol} reduces to
\begin{gather}
\frac{N_3(z)}{\mathcal N}=J_r\big(e^{-\Gamma(z-z_0)\sgn\eta}\big),\label{eq:J}\\
J_r(K)=\int_0^\infty\frac{dy}{\big(1+y+Ky^r\big)^2},\label{eq:Jdef}\\
\Gamma=2|\eta|w_1w_2=\frac{2r(1-r)W}{T_3},\label{eq:Gamma}
\end{gather}
and $N_{1,2}=\mathcal N-N_3$ by Manley--Rowe, where $T_3=1/(|\eta|W)$ is the pump duration.
The rate $\Gamma$ is the natural unit of depth for everything that follows: it is the inverse length over which one event converts its photons. In the optical variables
\begin{equation}
\Gamma=\frac{2\,|\beta_{1,3}-\beta_{1,1}|\,|\beta_{1,3}-\beta_{1,2}|}{|\beta_{1,1}-\beta_{1,2}|\,T_3},
\end{equation}
i.e., $\Gamma^{-1}$ is of the order of the walk-off length of the pump pulse with respect to the subharmonic waves.

The function $J_r$ depends only on $r$ and is symmetric under $r\to1-r$; the pole enters only through $\eta$, in the rate $\Gamma$. One has $J_r(0)=1$, complete conversion, and $J_r(\infty)=0$; at the vertex, $K=1$, the converted fraction lies between $1/2$ ($r\to0$) and $\tfrac43\big(1-\tfrac{\pi}{3\sqrt3}\big)\simeq0.527$ ($r=1/2$), so that the vertex is the point where about half of the photons have been converted. For $r\to0$ and $r=1/2$ the function is elementary,
\begin{align}
    J_0(K)&=\frac{1}{1+K}, \\
    J_{1/2}(K)&=\frac{4}{4-K^2}\left[1-\frac{K\arccos(K/2)}{\sqrt{4-K^2}}\right],
\end{align}
the logistic function and a curve close to it [Fig.~\ref{fig:faces}(b)]. For $r>0$ the law is no longer logistic, although it stays close to it, $|J_r(K)-1/(1+K)|\le0.037$ [Fig.~\ref{fig:faces}(b)].

\subsection{Where the conversion takes place}

The three legs are centered on the lines $\eta w_i(x-v_it)+\ln|n^0_j/n^0_k|=0$, which are concurrent because $\sum_iw_i=\sum_iw_iv_i=0$ and $\sum_i\ln|n^0_j/n^0_k|=0$. Their common point, where the three $|n_j|$ are equal, is
\begin{equation}\label{eq:vertex}
x_0=\frac{\sum_jv_jw_j\ln|n^0_j|}{\eta\,w_1w_2w_3},\qquad t_0=\frac{\sum_jw_j\ln|n^0_j|}{\eta\,w_1w_2w_3},
\end{equation}
(from $\sum_jw_jB_{jj}=\sum_jw_jv_j^2=-w_1w_2w_3$). It depends only on the moduli $|n^0_j|$, two real parameters since $\bm n^0$ is defined up to a factor, and scales as $1/\eta$; $\xi$ and the phases of $\bm n^0$ fix only the phases of the fields. In the parametrization \eqref{eq:physpar}, $(x_0,t_0)=(x_k,t_k)$.

For a single fusion event the depth of the vertex is purely kinematic. If the input signal and idler pulses are centered at $\tau_1^{\rm in}$ and $\tau_2^{\rm in}$ on the input face, the legs $\tau=\tau_{1,2}^{\rm in}+v_{1,2}z$ meet at
\begin{equation}\label{eq:z0fusion}
z_0=\frac{\tau_1^{\rm in}-\tau_2^{\rm in}}{W},\qquad \tau_0=\tau_1^{\rm in}+v_1z_0 ,
\end{equation}
independently of $g$ and $\eta$. What the soliton adds to kinematics is (i) the constraint that the two input pulses have the durations $T_{1,2}=1/(\eta w_{1,2})$, i.e., $w_1T_1=w_2T_2$, and areas $\pi$ in their blocks, and (ii) the conversion law \eqref{eq:J} around $z_0$. For a decay event, instead, the vertex is not fixed by the pump alone: the pump leg is a $\mathrm{sech}$ pulse for every value of $\ln|n^0_3|$, which enters only through the exponentially small signal and idler tails, and the depth of the decay is set by those tails, i.e., by a seed; the delay of the decay grows logarithmically as the seed vanishes \cite{KRB79}. In both cases the dependence of the vertex on the other solitons is the subject of Sec.~\ref{sec:shifts} below.

\subsection{Input/output and minimal crystal length}\label{sec:faces}

\begin{figure*}[t]
\includegraphics[width=\textwidth]{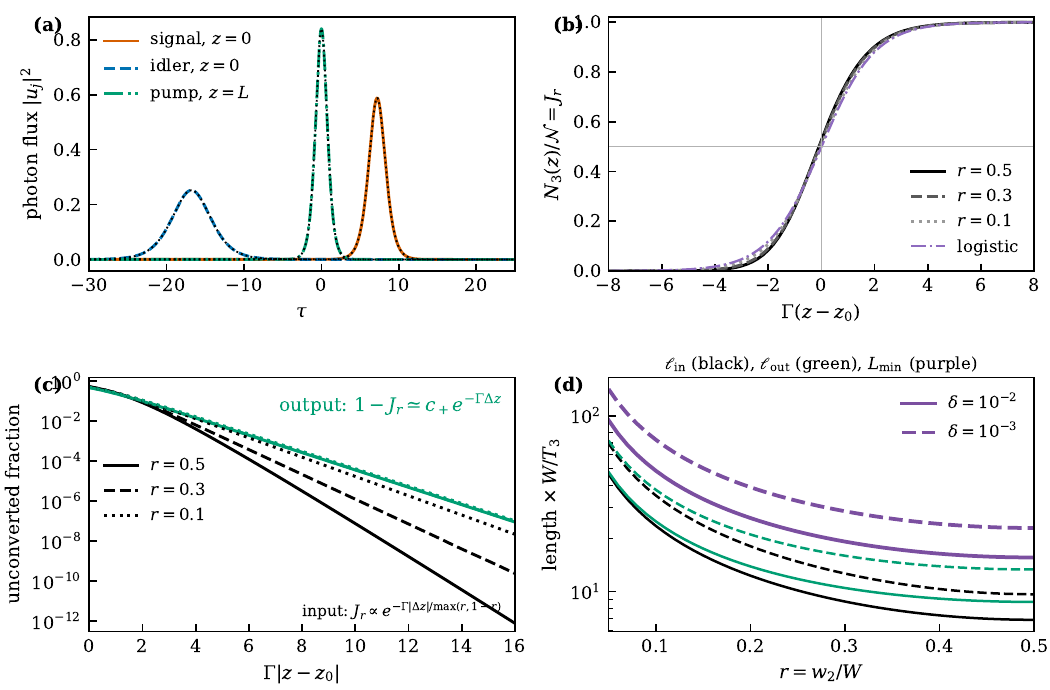}
\caption{\label{fig:faces}(a) Fluxes on the faces for the fusion event of Fig.~\ref{fig:events}: signal and idler on the input face, pump on the output face; dotted: asymptotic legs, Eq.~\eqref{eq:leg}. (b) Conversion curve $J_r$, Eq.~\eqref{eq:J}, for three values of $r$, compared with a logistic function of the same rate. (c) Unconverted fraction on the input side ($J_r$, black) and on the output side ($1-J_r$, green) on a logarithmic scale; the output side decays at rate $\Gamma$ for every $r$, the input side at rate $\Gamma/\max(r,1-r)$. (d) Minimum distances $\ell_{\rm in}$ (black), $\ell_{\rm out}$ (green) of the vertex from the input and output faces and minimum crystal length $L_{\min}=\ell_{\rm in}+\ell_{\rm out}$ (purple) for a fusion event, in units of $T_3/W$, for tolerances $\delta=10^{-2}$ (solid) and $10^{-3}$ (dashed) on the photon numbers; for a decay event $\ell_{\rm in}$ and $\ell_{\rm out}$ are exchanged.}
\end{figure*}

Far from the vertex the fields on the faces are superpositions of the $\mathrm{sech}^2$ legs \eqref{eq:leg}. In the optical variables a leg of wave $i$ has duration $T_i$, peak intensity
\begin{gather}
I_i^{\rm pk}=\frac{\hbar\omega_i\,|w_jw_k|}{\gamma^2T_i^2}=\frac{\varepsilon_0c^3n_1n_2n_3\,|w_jw_k|}{2d_{\rm eff}^2\,\omega_j\omega_k\,T_i^2},\label{eq:Ipk}
\end{gather}
and photon fluence $\mathcal N$, the photon number per unit area of the one-dimensional model. For a fusion event the input face shows the signal and idler pulses at $\tau_{1,2}^{\rm in}$ and the output face the pump pulse at $\tau_3^{\rm out}=\tau_0+v_3(L-z_0)$ [Fig.~\ref{fig:faces}(a)]; for a decay event the roles are exchanged. The face timings determine the vertex, Eq.~\eqref{eq:z0fusion}, so the depth of the conversion is measurable from outside the crystal.

The asymptotic description of the faces is accurate when the vertex is far from both of them, and the conversion curve \eqref{eq:J} applies. We require that on the input face the field contains a fraction smaller than $\delta$ of converted photons, and that on the output face a fraction smaller than $\delta$ is still unconverted. From the asymptotics of $J_r$,
\begin{gather}
1-J_r(K)\simeq c_+K\quad(K\to0),\qquad c_+=\frac{\pi r(1-r)}{\sin\pi r},\\
J_r(K)\simeq C_rK^{-1/(1-r)}\quad(K\to\infty,\ r<\tfrac12),\\
C_r=\frac{\pi r}{(1-r)^2\sin\frac{\pi(1-2r)}{1-r}},
\end{gather}
($J_{1/2}\simeq4K^{-2}\ln K$), the vertex of a fusion event must lie at depth $z_0\ge\ell_{\rm in}$ and the crystal must extend beyond it by $\ell_{\rm out}$, with
\begin{equation}\label{eq:lengths}
\ell_{\rm in}\simeq\frac{\max(r,1-r)}{\Gamma}\ln\frac{C_r}{\delta},\qquad
\ell_{\rm out}\simeq\frac{1}{\Gamma}\ln\frac{c_+}{\delta},
\end{equation}
and the minimum crystal length is $L_{\min}=\ell_{\rm in}+\ell_{\rm out}$. For a decay event $\ell_{\rm in}$ and $\ell_{\rm out}$ are exchanged ($z\to-z$, $\eta\to-\eta$). We use a photon-number criterion because it refers directly to the pulse energies measured on the faces. The values obtained by inverting $J_r$ are shown in Fig.~\ref{fig:faces}(d). For $r=1/2$,
\begin{equation}\label{eq:Lmin}
\begin{aligned}
L_{\min}&=15.6\,T_3/W\qquad(\delta=10^{-2}),\\
L_{\min}&=23.0\,T_3/W\qquad(\delta=10^{-3}),
\end{aligned}
\end{equation}
with $\ell_{\rm in}=6.9\,T_3/W$ and $\ell_{\rm out}=8.7\,T_3/W$ for $\delta=10^{-2}$, and, at fixed pump duration, $L_{\min}$ grows as $1/[r(1-r)]$. Hence $r=1/2$ is the optimal choice, and when the pump group velocity approaches that of the signal or of the idler ($r\to0$ or $r\to1$) a shorter pump pulse is needed for the faces of a given crystal to carry the legs. Physical values are given in Sec.~\ref{sec:scales}.

\subsection{Events and their asymptotic profiles}\label{sec:profiles}

Beyond the lengths \eqref{eq:lengths} the faces carry the asymptotic legs \eqref{eq:leg}, and it is natural to describe the output in terms of these pulses alone. Three remarks show what this description retains and what it leaves out.

\emph{The legs are not self-sustained.} On the ray $x\simeq v_it$ the other two fields vanish exponentially, and Eq.~\eqref{eq:3wi} reduces to $(\partial_t+v_i\partial_x)u_i=0$: any profile of a single wave propagates unchanged. The $\mathrm{sech}$ shape and the area $\pi$ of a leg are not maintained by a balance between nonlinearity and dispersion, as for a soliton of the nonlinear Schr\"odinger equation; they are fixed at the vertex, where the three waves overlap. The legs have the profile of Zakharov--Shabat solitons because, where only wave $i$ is present, the $3\times3$ problem reduces to the Zakharov--Shabat problem of the block $(j,k)$ (Sec.~\ref{sec:inputface}).

\emph{One event, three spectra.} For separated pulses the scattering data split into three Zakharov--Shabat spectra, one for each wave \cite{KRB79,Ibragimov99}. An event is a single pole whose vector $\bm n^0$ has three nonzero components. In the factorized description it appears as two coinciding eigenvalues, of the signal and of the idler, on one face and as one pump eigenvalue on the other. That these are the same object, with a common $\eta$, a common vertex and the locked phase $\arg u_3=\arg(u_1u_2)+{\rm const}$, is a property of the $3\times3$ problem.

The same holds for the shape of several solitons of one wave. A Zakharov--Shabat $n$-soliton is fixed by its eigenvalues and by its norming constants, which in the present parametrization are the moduli and phases of the vectors $\bm n^0_k$. The bell-shaped profile $n\,\mathrm{sech}$ used in Ref.~\cite{Ibragimov99} is a particular point of these data: eigenvalues in the ratio $1:3:\dots:(2n-1)$ \cite{SY74}, and norming constants that place all solitons at the same point with alternating signs, for instance $\bm n^0_k=(1,(-1)^k,0)$ and $|\eta_k|=(2n-1-2k)/(WT)$ for a pump pulse of duration $T$. It is the natural form of a smooth pulse on the input face. On the output face the legs of one wave travel at a common velocity and keep the relative positions set by their vertices and by the shifts of Sec.~\ref{sec:shifts}; they recombine into a symmetric profile only for specially prepared inputs, such as the shaped trigger pulses of Ref.~\cite{Ibragimov99}. The general output is a network of events with arbitrary $\bm n^0_k$.

\emph{Correlated legs.} The three legs of an event pass through its vertex, so that whatever moves the vertex moves all of them. For a decay event the pump leg is fixed and, by Eq.~\eqref{eq:vertex}, a change of $\ln|n^0_3|$ displaces the vertex along the pump characteristic, $\delta x_0=v_3\,\delta t_0$, with $\delta t_0=\delta\ln|n^0_3|/(\eta w_1w_2)$. The outgoing legs $x=x_0+v_i(t-t_0)$ are then displaced on the output face by $\delta\tau_i=(v_3-v_i)\,\delta z_0$, i.e.,
\begin{equation}\label{eq:corr}
\delta\tau_1=w_2\,\delta z_0,\qquad \delta\tau_2=-w_1\,\delta z_0,\qquad \frac{\delta\tau_1}{\delta\tau_2}=-\frac{r}{1-r},
\end{equation}
whatever the cause of the displacement and the value of $\eta$. When the depth of a decay is set by a seed, fluctuations of the seed move the signal and idler output times along a fixed line, with opposite signs relative to the pump. Conversely, for a fusion event Eq.~\eqref{eq:z0fusion} gives the arrival time of the pump on the output face as the weighted mean $\tau_0=(1-r)\tau_1^{\rm in}+r\,\tau_2^{\rm in}$ of the input times, in the pump frame. For parametric amplification this correlation is contained in the output timings of Ref.~\cite{Ibragimov99} [Eqs.~(20) and (21) there], in which the reflection coefficient of the trigger pulse enters the times of both outputs; here it is a kinematic consequence of the common vertex, and holds whatever displaces it.

\subsection{Two solitons}\label{sec:two}

For $N=2$, with $\mu_k=\xi_k+i\eta_k$, $\mu_1\neq\mu_2$, $\mu_2\neq\bar\mu_1$, $n_{k,j}=n^0_{k,j}e^{-i\mu_k\zeta_j}$, let
\begin{gather}
D_k=\sum_js_j|n_{k,j}|^2,\qquad p=\bm n_1^\dagger S\,\bm n_2,\\
\theta=\frac{\mu_1-\mu_2}{\mu_1-\bar\mu_2},\qquad 1-|\theta|^2=\frac{4\eta_1\eta_2}{|\mu_1-\bar\mu_2|^2}.\label{eq:theta}
\end{gather}
Then $\mathcal D\equiv-4\eta_1\eta_2\det M=D_1D_2-(1-|\theta|^2)|p|^2$ and $Q_{jk}=2(v_j-v_k)\,s_k\,\mathcal N_{jk}/\mathcal D$, with
\begin{multline}
\mathcal N_{jk}=\eta_1D_2\,n_{1,j}\bar n_{1,k}+\eta_2D_1\,n_{2,j}\bar n_{2,k}\\
-2i\eta_1\eta_2\Big[\frac{p\,n_{1,j}\bar n_{2,k}}{\mu_2-\bar\mu_1}+\frac{\bar p\,n_{2,j}\bar n_{1,k}}{\mu_1-\bar\mu_2}\Big],
\end{multline}
so that
\begin{equation}\label{eq:2sol}
\begin{aligned}
u_1&=\frac{2}{g}\,w_1\sqrt{|w_2w_3|}\;\frac{s_3\,\mathcal N_{23}}{\mathcal D},\\
u_2&=\frac{2}{g}\,w_2\sqrt{|w_3w_1|}\;\frac{s_1\,\mathcal N_{31}}{\mathcal D},\\
u_3&=-\frac{2i}{g}\,|w_3|\sqrt{|w_1w_2|}\;\frac{s_1\,\mathcal N_{21}}{\mathcal D}.
\end{aligned}
\end{equation}
For a single pole $\mathcal N_{jk}\to\eta D\,n_j\bar n_k$, $\mathcal D\to D^2$, and Eq.~\eqref{eq:1sol} is recovered. The solution is regular for $S=\pm\mathbb I$: if $\eta_1\eta_2<0$ then $1-|\theta|^2<0$ and $\mathcal D>D_1D_2>0$; if $\eta_1\eta_2>0$ then $|p|^2\le D_1D_2$ and $\mathcal D\ge|\theta|^2D_1D_2>0$.

\emph{Degenerate polarization.} The dressing factorizes into one \emph{Darboux factor} per pole, $\chi=\chi_N\cdots\chi_1$, with
\begin{gather}
\chi_k(\lambda)=\mathbb I-\frac{\mu_k-\bar\mu_k}{\lambda-\bar\mu_k}\,\frac{\tilde{\bm n}_k\tilde{\bm n}_k^\dagger S}{\tilde{\bm n}_k^\dagger S\,\tilde{\bm n}_k},\label{eq:darboux}\\
\tilde{\bm n}_k=\chi_{k-1}\cdots\chi_1(\mu_k)\,\bm n_k,\nonumber
\end{gather}
i.e., each factor is built on the vector of its pole dressed by the factors already applied; $\chi_k(\mu_k)\tilde{\bm n}_k=0$, the product does not depend on the order of the poles \cite{ZS79}, and for $N=1$ it is Eq.~\eqref{eq:gram}. Let $\bm e_m$ be the $m$th unit vector of $\mathbb C^3$ and $E_{mm}=\bm e_m\bm e_m^T$. If $\bm n^0_2=c\,\bm e_m$, then $\bm n_2\propto\bm e_m$ everywhere, since $Z$ is diagonal, and with pole 2 taken first its factor is diagonal, $\chi_2(\lambda)=\mathbb I+\big(\tfrac{\lambda-\mu_2}{\lambda-\bar\mu_2}-1\big)E_{mm}$. It commutes with $A$, so that it does not contribute to $Q$, and it dresses the vector of pole 1 into $\chi_2(\mu_1)\bm n_1$, which differs from $\bm n_1$ only in the $m$th component, multiplied by $(\mu_1-\mu_2)/(\mu_1-\bar\mu_2)=\theta$. Hence Eq.~\eqref{eq:2sol} reduces exactly to Eq.~\eqref{eq:1sol} with
\begin{equation}\label{eq:lemma}
n^0_{1,m}\to\theta\,n^0_{1,m},\qquad n^0_{1,j\neq m}\ \text{unchanged}.
\end{equation}

\begin{figure*}[t]
\includegraphics[width=\textwidth]{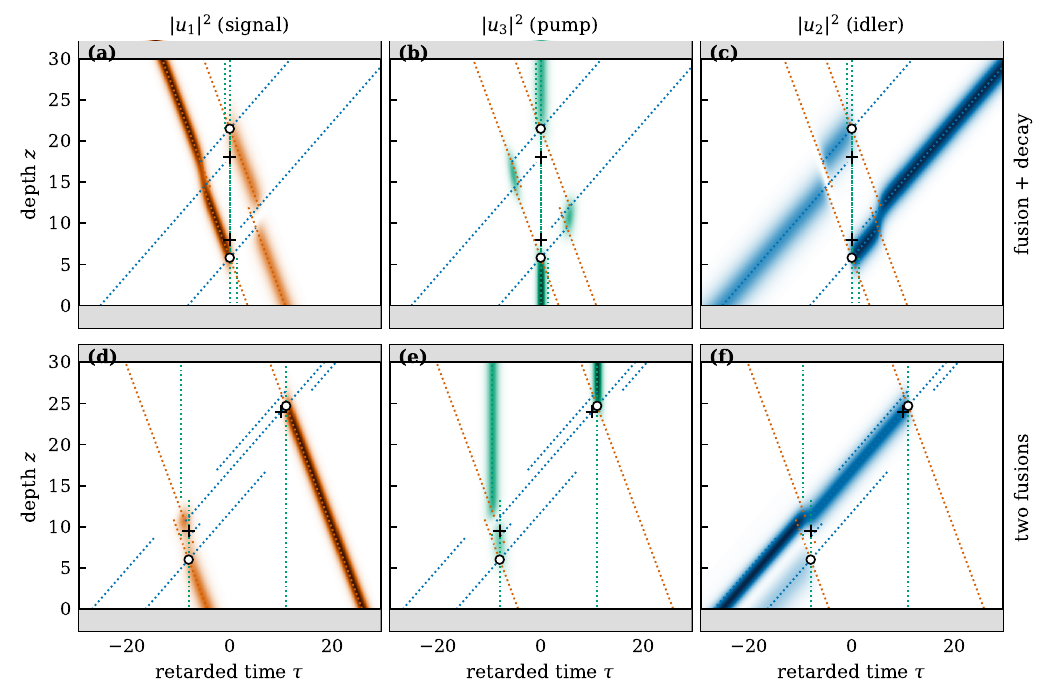}
\caption{\label{fig:two}Two-soliton solutions, $r=0.3$, $L=30$. Crosses: nominal vertices $(\tau_k,z_k)$ of Eq.~\eqref{eq:physpar}, i.e., the vertices that each event would have without the other; open circles: actual vertices, given by the shift rule, Eq.~\eqref{eq:shift}; dotted lines: centers of the legs (orange: signal, blue: idler, green: pump) given by the dressed vectors \eqref{eq:dressed}, which follow the characteristics through the actual vertices and jump where they cross a leg of the other event. (a)--(c) A fusion ($\mu_1=0.5i$) above a decay ($\mu_2=-0.8i$) with aligned pumps: the products of the decay cross the incoming signal and idler of the fusion. At each of the two crossings a signal and an idler from different events fuse into a short-lived pump pulse [the isolated segments in (b)] and separate again. The vertices are pushed apart along the pump characteristic, from the nominal depths $8$ and $18$ to $5.8$ and $21.5$, while far from the crossings the legs lie where the two one-event solutions would put them. (d)--(f) Two fusions ($\mu_1=0.5i$, $\mu_2=0.15+0.75i$): the incoming idler of the second event crosses the outgoing pump of the first, generating a transient signal [the isolated segment above the lower vertex in (d)], and jumps; the vertices are displaced from their nominal depths $9.5$ and $24$ to $6.0$ and $24.7$.}
\end{figure*}

\emph{Asymptotics.} Let $(i,j,l)$ be cyclic and $\sigma_i=\sgn[(v_i-v_j)(v_i-v_l)]$ ($+1$ for the extreme velocities, $-1$ for the intermediate one). On the ray $x=v_it+\tilde x$, $\tilde x=O(1)$,
\begin{equation}
\zeta_j-\zeta_l=w_i\,\tilde x,\qquad \zeta_i-\zeta_j\simeq(v_i-v_j)(v_i-v_l)\,t,
\end{equation}
so for $|t|\to\infty$ the vector $\bm n_k$ is either dominated by its $i$ component, $\bm n_k\propto\bm e_i$ (if $\eta_k\sigma_it>0$), or has a negligible $i$ component (if $\eta_k\sigma_it<0$). In the first case pole $k$ has no leg on the ray and, by Eq.~\eqref{eq:lemma}, only rescales the $i$ component of the other pole, which is negligible there: it has no effect at all. Pole $k$ thus contributes a leg on the ray $x\simeq v_it$ on the side $\eta_k\sigma_it<0$ only, exactly as for one pole. Two cases arise.
\begin{itemize}
\item[(a)] $\eta_1\eta_2<0$ (opposite orientations). On every ray and on each side exactly one pole is present, and its leg is the one-pole $\mathrm{sech}$ \emph{without any shift}:
\begin{multline}
|u_i|\to\frac{|\eta_k w_i|\sqrt{|w_jw_l|}}{g}\\
\times\mathrm{sech}\big[\eta_kw_i(x-v_it)+\ln|n^0_{k,j}/n^0_{k,l}|\big],
\end{multline}
$k$ being the pole with $\eta_k\sigma_it<0$. For $\eta_1>0>\eta_2$ the input is $u_1,u_2$ from pole 1 and $u_3$ from pole 2, the output $u_3$ from pole 1 and $u_1,u_2$ from pole 2: a fusion and a decay whose legs cross, leaving all asymptotic positions and amplitudes unchanged [Fig.~\ref{fig:two}(a)--(c)]. Between the crossings, and in particular at the vertices, the legs are displaced; the displacements are undone where the legs cross.
\item[(b)] $\eta_1\eta_2>0$ (same orientation). Both poles have their legs on the same side of the same rays (for $\eta_{1,2}>0$: on the rays $v_{1,2}$ for $t\to-\infty$, on $v_3$ for $t\to+\infty$). Two legs on the same ray do not separate: there $u_i$ is the two-soliton of the Zakharov--Shabat problem of the $(j,l)$ block, which beats at frequency $(\xi_1-\xi_2)w_i$ when $\xi_1\neq\xi_2$ and carries $N_i=2(|\eta_1|+|\eta_2|)|w_1w_2w_3|/g^2$ photons. When the two centers $\tilde x_k=-\ln|n^0_{k,j}/n^0_{k,l}|/(\eta_kw_i)$ are far apart, Eq.~\eqref{eq:lemma} applied inside the block gives two $\mathrm{sech}$ pulses travelling together,
\begin{multline}
|u_i|\to\sum_{k=1,2}\frac{|\eta_k w_i|\sqrt{|w_jw_l|}}{g}\\
\times\mathrm{sech}\big[\eta_kw_i(x-v_it)+\ln|n^0_{k,j}/n^0_{k,l}|+\nu_k\ln|\theta|\big],
\end{multline}
$\nu_1=\sgn[\eta_2w_i(\tilde x_1-\tilde x_2)]$, $\nu_2=\sgn[\eta_1w_i(\tilde x_2-\tilde x_1)]$. Since $|\theta|<1$ the two pulses are pushed apart, their separation growing by $\ln|\theta|^{-1}\big(|\eta_1w_i|^{-1}+|\eta_2w_i|^{-1}\big)$. Depending on the vertices, the solution looks like a single Y with doubled legs or like two nested Y's of the same orientation [Fig.~\ref{fig:two}(d)--(f)].
\end{itemize}
In both cases $\mathcal Q_1$ and $\mathcal Q_2$ computed from the asymptotic legs coincide for $t\to\pm\infty$, as required by the Manley--Rowe conservation laws.

\emph{Reading the two-soliton solution.} Far from all crossings the solution is the superposition of two Y's, but two features distinguish it from a mere superposition (Fig.~\ref{fig:two}). First, each event takes place at an \emph{actual vertex} that differs from its \emph{nominal vertex}, the vertex that the same pole $\mu_k$ and vector $\bm n^0_k$ would have in the absence of the other pole. Second, where a leg of one event crosses a leg of the other, the photons are partly converted for a short distance: a signal and an idler from different events fuse into a short-lived pump pulse, and a signal (idler) crossing a pump generates a transient idler (signal). While converted, the photons travel at the group velocity of another wave, so that the leg leaves the crossing displaced. These displacements are discussed at length, for any number of events, in Sec.~\ref{sec:shifts}.

\subsection{$N$ solitons: networks of events and the shift rule}\label{sec:shifts}

The structure of the two-soliton solution extends to any $N$, and gives a direct physical reading of Eq.~\eqref{eq:Nsol}. Each pole $k$ is a conversion event, fusion or decay, whose Y divides the $(\tau,z)$ plane into three \emph{sectors}: sector $m$ is the wedge bounded by the two legs other than the leg of wave $m$, and in it the $m$th component of $\bm n_k$ dominates exponentially (for a fusion: sector 3 is the wedge between the incoming signal and idler, sectors 1 and 2 lie on either side of the outgoing pump). In sector $m$ the dressing factor of pole $k$ is diagonal up to exponentially small terms, and since the Darboux factors \eqref{eq:darboux} can be applied in any order, Eq.~\eqref{eq:lemma} generalizes as follows. Near an event $l$ located, for every $k\neq l$, in sector $m_k$ of event $k$, the $N$-soliton coincides with the one-soliton \eqref{eq:1sol} with dressed polarization
\begin{equation}\label{eq:dressed}
\tilde n^0_{l,j}=n^0_{l,j}\prod_{k\neq l:\ m_k=j}\theta_{lk},\qquad \theta_{lk}=\frac{\mu_l-\mu_k}{\mu_l-\bar\mu_k}.
\end{equation}
By Eq.~\eqref{eq:vertex} the vertex of event $l$ is displaced from its nominal position by
\begin{equation}\label{eq:shift}
\Delta z_l=\sum_{k\neq l}\frac{w_{m_k}\ln|\theta_{lk}|}{\eta_l\,w_1w_2w_3},\qquad \Delta\tau_l=\sum_{k\neq l}v_{m_k}\frac{w_{m_k}\ln|\theta_{lk}|}{\eta_l\,w_1w_2w_3},
\end{equation}
and likewise every leg of event $l$ is translated in $\tau$ by $\Delta\tau=-\big[\ln|\tilde n^0_{l,j}/\tilde n^0_{l,k}|-\ln|n^0_{l,j}/n^0_{l,k}|\big]/(\eta_lw_i)$ wherever the other events are in definite sectors. The corrections are exponentially small in the distance from the legs of the other events, since such are the off-diagonal parts of their dressing factors. When a leg of event $l$ crosses a leg of event $k$, it passes from one sector of $k$ to another, the component multiplied by $\theta_{lk}$ changes, and the leg jumps; physically, the photons that are converted for a short distance at the crossing travel meanwhile at another group velocity, as in the resonance of Sec.~\ref{sec:resonance}. The nominal vertices depend on the parametrization, since $\bm n^0_l$ can be redefined to absorb the factors \eqref{eq:dressed} in one sector; the jumps between sectors, $w_{m}\ln|\theta_{lk}|$, and the positions on the faces do not.

Equation~\eqref{eq:shift} has three features worth stressing. First, the contribution of each other event displaces the vertex \emph{along the characteristic of wave $m_k$}, $\Delta\tau=v_{m_k}\Delta z$: the event is advanced or delayed as if it had been carried by the wave whose leg faces the sector. Second, the contributions are pairwise and additive, a direct consequence of integrability: there are no three-body effects, and the network of $N$ events is fixed by the $N$ one-event data and by the $N(N-1)/2$ numbers $|\theta_{lk}|$. Third, the size of the shift is set by the spectral distance of the events, through $|\theta_{lk}|$, and its sign by their relative orientation: $\ln|\theta_{lk}|<0$ for events of the same orientation and $>0$ for opposite orientations, so that, with $w_1w_2w_3<0$, a fusion event in sector 1 or 2 of another fusion is delayed (pushed deeper into the crystal) and one in sector 3 is advanced. Events with very different detunings, $|\xi_l-\xi_k|\gg|\eta_l|,|\eta_k|$, do not shift each other, $\ln|\theta_{lk}|\simeq-2\eta_l\eta_k/(\xi_l-\xi_k)^2$.

\begin{figure*}[t]
\includegraphics[width=\textwidth]{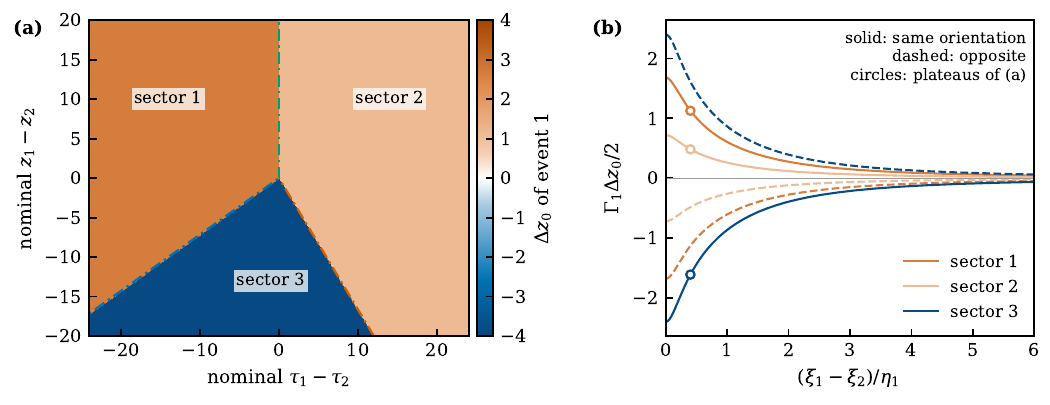}
\caption{\label{fig:shift}(a) Depth shift $\Delta z_0$ of a fusion event ($\mu_1=0.2+0.5i$), Eq.~\eqref{eq:shift}, as a function of its nominal position relative to the vertex of a second fusion event ($\mu_2=0.6i$), $r=0.3$; lines: legs of the second event, which bound its sectors. The plateaus are $\Delta z_0=2.68$, $1.15$ and $-3.83$ in sectors 1, 2 and 3, as given by Eq.~\eqref{eq:shift}. (b) Depth shift in units of $2/\Gamma_1$, with $\Gamma_1=2|\eta_1|w_1w_2$ the rate of the first event, $\Gamma_1\Delta z_0/2=(w_m/w_3)\ln|\theta_{12}|$ in sector $m$, as a function of the detuning $\xi_1-\xi_2$, for $\eta_1=0.5$ and $\eta_2=0.6$ (solid: same orientation) or $\eta_2=-0.6$ (dashed: opposite orientation). Since $\eta_2\to-\eta_2$ maps $|\theta_{12}|\to1/|\theta_{12}|$, the dashed curves are the mirror images of the solid ones. Open circles: the plateaus of (a), at $(\xi_1-\xi_2)/\eta_1=0.4$. Colors as the sectors in (a).}
\end{figure*}

Near the legs of the other events the dressing factors are not diagonal and Eq.~\eqref{eq:shift} does not apply: within a few pulse durations of a leg the vertex is not sharply defined, and the corrections to Eq.~\eqref{eq:shift} decay exponentially with the distance from it. Figure~\ref{fig:shift}(a) shows the shift of a fusion event as its nominal position is moved across the Y of a second fusion: it is piecewise constant, with the three values of Eq.~\eqref{eq:shift}, and changes where the nominal position crosses a leg of the other event. The three plateaus are in the ratio $w_1:w_2:w_3$ and sum to zero, so that the largest shift, opposite to the other two, is in sector 3, between the incoming signal and idler. Figure~\ref{fig:shift}(b) shows the dependence on the spectral parameters: the shift is largest for equal detunings, decays as $(\xi_1-\xi_2)^{-2}$ for detunings large compared with $\eta_{1,2}$, and changes sign with the relative orientation of the two events.

\begin{figure*}[t]
\includegraphics[width=\textwidth]{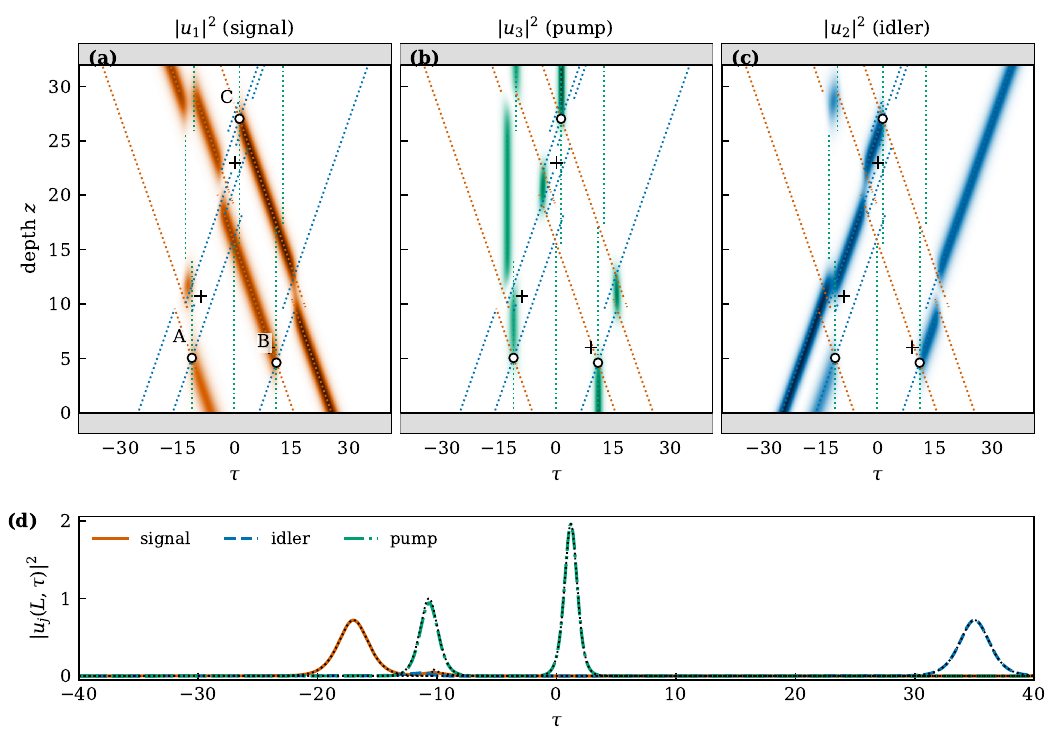}
\caption{\label{fig:three}Network of three events, $r=1/2$, $L=32$: fusion A ($\mu=0.5i$), decay B ($\mu=-0.05-0.6i$), fusion C ($\mu=0.1+0.7i$). Crosses: nominal vertices; open circles: actual vertices, Eq.~\eqref{eq:shift}; dotted lines: centers of the legs, Eq.~\eqref{eq:dressed}, which jump at each crossing with a leg of another event. (a)--(c) Fluxes in the crystal; the isolated pump segments in (b) are the short-lived pump pulses generated where a signal and an idler of different events cross. (d) Fluxes on the output face (colored) and prediction of the shift rule, Eq.~\eqref{eq:dressed} (dotted).}
\end{figure*}

\emph{An example.} Figure~\ref{fig:three} shows a network of three events: a fusion A and a decay B at the same depth, and a fusion C deeper in the crystal whose incoming idler crosses the pump of A and the outgoing signal of B, and whose incoming signal crosses the outgoing idler of B; the outgoing signal of B also crosses the pump of A. At each crossing of a signal and an idler leg a pump pulse is generated and reabsorbed (the short pump segments in Fig.~\ref{fig:three}(b), resonances of Sec.~\ref{sec:resonance}); the only lasting effect of the crossings is the displacement of the legs. The output face carries four pulses, the pumps of A and C and the signal and idler of B, and their positions and shapes are those predicted by Eq.~\eqref{eq:dressed}, up to corrections exponentially small in the distance of the last crossing from the face. The Manley--Rowe charges $Q_{1,2}$ are, as expected, the same on the two faces.

\subsection{Input pulse spectra}\label{sec:inputface}

In the propagation reading, the field on the input face is the initial datum, and the inverse scattering transform maps it to the field on the output face for any crystal length. The scattering data split into a discrete part, the poles, and a continuous part, the radiation. Since the model has no dispersion, radiation in wave $j$ travels asymptotically at $v_j$ without spreading, while the poles produce the conversion events and the pure pulses of Sec.~\ref{sec:pure}. For a purely solitonic input, in a crystal long compared with the lengths of Sec.~\ref{sec:faces}, the output face is a linear superposition of $\mathrm{sech}$ legs, nonlinear only where legs overlap, and its only lasting trace is the set of shifts \eqref{eq:shift}.

When the input pulses do not overlap, as is the case for walk-off-separated pulses on the input face, the scattering matrix factorizes into those of the individual pulses \cite{KRB79,Ibragimov99}. A pulse of wave $m$ alone is a potential $q_m=\beta_mu_m$ of the $2\times2$ focusing Zakharov--Shabat problem of the block complementary to $m$, i.e., of the restriction of the $3\times3$ problem to the two components $j,l\neq m$, in which $q_m$ is the only off-diagonal entry (focusing because $\rho_m=-1$ for all $m$ when $S=\mathbb I$), with spectral parameter $\zeta=\lambda w_m/2$. Its discrete eigenvalues $\zeta_n$ give poles of the three-wave problem \cite{KRB79} at
\begin{equation}\label{eq:eigmap}
\mu_n=\frac{2\zeta_n}{w_m},\qquad |\eta_n|=\frac{2\,\Im\zeta_n}{|w_m|}.
\end{equation}
For a $\mathrm{sech}$ pulse of duration $T$ and area $h\pi$ in its block, $\beta_m\int|u_m|d\tau=h\pi$, the eigenvalues are $\zeta_n=i(h-\tfrac12-n)/T$, $0\le n<h-\tfrac12$ \cite{SY74}, so that
\begin{equation}\label{eq:etasech}
|\eta_n|=\frac{2h-1-2n}{|w_m|\,T}.
\end{equation}
A single-humped real pulse carries no soliton for $h\le1/2$, which is the area threshold of the three-wave interaction \cite{KS03,KRB79}. The area, rather than the energy, is the relevant quantity because the discrete spectrum of the Zakharov--Shabat problem is controlled by $\int|q_m|d\tau$: there are no eigenvalues for $\int|q_m|d\tau<\pi/2$ \cite{KS03}, the eigenvalues of a real single-humped pulse are purely imaginary \cite{KS02}, and for such pulses the number of solitons is, to a good approximation, the integer closest to the normalized area \cite{KRB79,Ibragimov99}; for the $\mathrm{sech}$ profile this count is exact, Eq.~\eqref{eq:etasech}. In the optical variables $h=\gamma\int|a_m|d\tau\,/\,(\pi\sqrt{|w_jw_l|})$: the threshold depends on the coupling and on the two group-velocity mismatches of wave $m$ with the other two waves. These statements assume an unchirped pulse, i.e., a real potential up to a constant phase. A linear phase, $u_m\propto e^{-i\Omega\tau}$, only shifts all eigenvalues by the same real amount, i.e., it changes the detuning $\xi$ and not $\eta$; a chirp, a phase that varies nonlinearly across the pulse, makes the potential genuinely complex, and the eigenvalues can then acquire different real parts, move toward the real axis or disappear into the continuum, so that a chirped pulse of the same area carries fewer events. We restrict ourselves to unchirped pulses. Two pulses of different waves whose potentials coincide up to the scaling $\tau\to\tau/(|w_m|T)$ have the same poles $\mu_n$, so that the matching discussed below is preserved if the chirps coincide in the scaled variable.

Two separated pulses of the same wave combine into a single potential of the same $2\times2$ block: the relevant spectral data are those of the whole block, not of the individual pulses (see Sec.~\ref{sec:resonance}).

\subsection{Pure pulses and the instability of the pump pulse}\label{sec:pure}

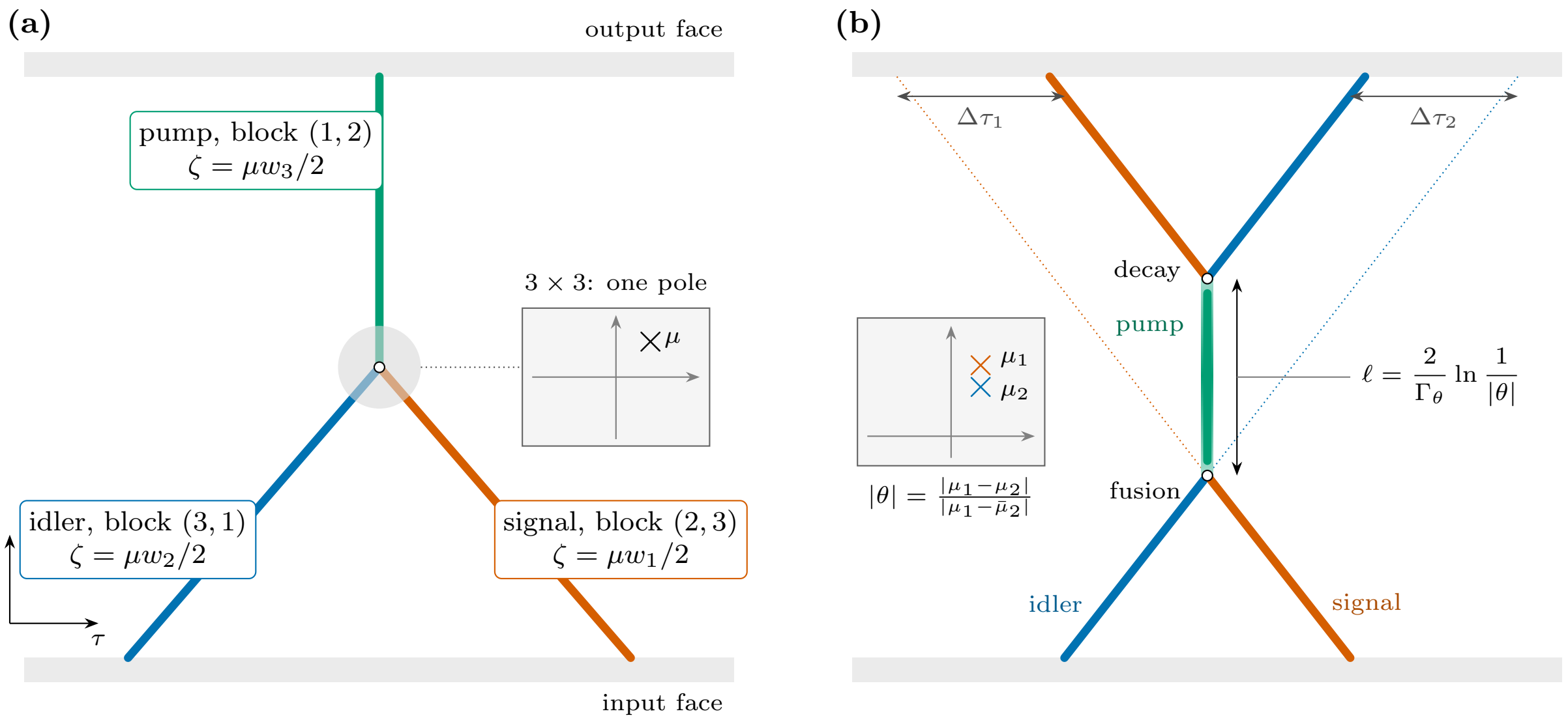
\begin{figure*}[t]
\centering
% Schematic of a conversion event and of a resonance (TikZ). Requires \usepackage{tikz}
% and \usetikzlibrary{arrows.meta,calc,decorations.pathreplacing}.
\definecolor{csig}{RGB}{213,94,0}
\definecolor{cidl}{RGB}{0,114,178}
\definecolor{cpmp}{RGB}{0,158,115}
\begin{tikzpicture}[x=1cm,y=1cm,font=\small,
  leg/.style={line width=2.4pt,line cap=round},
  blk/.style={draw,rounded corners=2pt,fill=white,inner sep=2.5pt,align=center,font=\footnotesize},
  pl/.style={draw=black!60,fill=black!4},
  >={Stealth[length=4pt]}]
% ------------------------------------------------------------------ (a)
\begin{scope}
  \node[anchor=north west,font=\bfseries\small] at (-3.9,3.75) {(a)};
  % faces
  \fill[black!8] (-3.6,-3.2) rectangle (3.6,-2.95);
  \fill[black!8] (-3.6,3.2) rectangle (3.6,2.95);
  \node[font=\scriptsize,anchor=east] at (3.6,-3.42) {input face};
  \node[font=\scriptsize,anchor=east] at (3.6,3.42) {output face};
  % legs
  \draw[leg,cidl] (-2.55,-2.95) -- (0,0);
  \draw[leg,csig] (2.55,-2.95) -- (0,0);
  \draw[leg,cpmp] (0,0) -- (0,2.95);
  % vertex: region where the three waves overlap
  \fill[black!15,opacity=.6] (0,0) circle (0.42);
  \fill[white] (0,0) circle (1.5pt); \draw (0,0) circle (1.5pt);
  % axes
  \draw[->] (-3.75,-2.6) -- (-3.75,-1.7) node[left,font=\scriptsize] {$z$};
  \draw[->] (-3.75,-2.6) -- (-2.85,-2.6) node[below,font=\scriptsize] {$\tau$};
  % spectral plane of the 3x3 problem
  \begin{scope}[shift={(1.45,-0.1)}]
    \draw[pl] (0,-0.7) rectangle (1.9,0.7);
    \draw[black!50,->] (0.1,0) -- (1.8,0);
    \draw[black!50,->] (0.95,-0.63) -- (0.95,0.63);
    \node[font=\large] at (1.3,0.36) {$\times$};
    \node[font=\scriptsize,right=-1pt] at (1.36,0.38) {$\mu$};
    \node[font=\scriptsize,anchor=south,align=center] at (0.95,0.72) {$3\times3$: one pole};
  \end{scope}
  \draw[densely dotted,black!70] (0.42,0) -- (1.45,0);
  % 2x2 blocks carried by the legs
  \node[blk,draw=cpmp] at (-1.25,2.2) {pump, block $(1,2)$\\ $\zeta=\mu w_3/2$};
  \node[blk,draw=csig] at (2.45,-1.75) {signal, block $(2,3)$\\ $\zeta=\mu w_1/2$};
  \node[blk,draw=cidl] at (-2.45,-1.75) {idler, block $(3,1)$\\ $\zeta=\mu w_2/2$};
\end{scope}
% ------------------------------------------------------------------ (b)
\begin{scope}[xshift=8.4cm]
  \node[anchor=north west,font=\bfseries\small] at (-3.9,3.75) {(b)};
  \fill[black!8] (-3.6,-3.2) rectangle (3.6,-2.95);
  \fill[black!8] (-3.6,3.2) rectangle (3.6,2.95);
  % free propagation (dotted) through the collision point (0,-1.1)
  \draw[densely dotted,cidl] (0,-1.1) -- (3.15,2.95);
  \draw[densely dotted,csig] (0,-1.1) -- (-3.15,2.95);
  % incoming pure pulses
  \draw[leg,cidl] (-1.45,-2.95) -- (0,-1.1);
  \draw[leg,csig] (1.45,-2.95) -- (0,-1.1);
  % pump resonance
  \shade[top color=cpmp!35,bottom color=cpmp!35,middle color=cpmp] (-0.06,-1.1) rectangle (0.06,0.9);
  \draw[leg,cpmp] (0,-0.95) -- (0,0.75);
  % outgoing legs, displaced
  \draw[leg,cidl] (0,0.9) -- (1.6,2.95);
  \draw[leg,csig] (0,0.9) -- (-1.6,2.95);
  \foreach \y in {-1.1,0.9} {\fill[white] (0,\y) circle (1.5pt); \draw (0,\y) circle (1.5pt);}
  \node[font=\scriptsize,anchor=east] at (-0.15,-1.25) {fusion};
  \node[font=\scriptsize,anchor=east] at (-0.15,0.98) {decay};
  \draw[<->] (0.3,-1.1) -- (0.3,0.9);
  \node[font=\scriptsize,anchor=west] at (1.45,-0.1) {$\ell=\dfrac{2}{\Gamma_\theta}\ln\dfrac{1}{|\theta|}$};
  \draw[black!50] (0.3,-0.1) -- (1.45,-0.1);
  % shifts on the output face
  \draw[<->,black!70] (-3.15,2.75) -- node[below,font=\scriptsize] {$\Delta\tau_1$} (-1.6+0.155,2.75);
  \draw[<->,black!70] (1.6-0.155,2.75) -- node[below,font=\scriptsize] {$\Delta\tau_2$} (3.15,2.75);
  % spectral plane: two nearby poles
  \begin{scope}[shift={(-3.55,-1.0)}]
    \draw[pl] (0,0) rectangle (1.9,1.5);
    \draw[black!50,->] (0.1,0.3) -- (1.8,0.3);
    \draw[black!50,->] (0.95,0.08) -- (0.95,1.42);
    \node[font=\large,csig] at (1.25,1.02) {$\times$};
    \node[font=\large,cidl] at (1.25,0.80) {$\times$};
    \node[font=\scriptsize,anchor=west] at (1.33,1.08) {$\mu_1$};
    \node[font=\scriptsize,anchor=west] at (1.33,0.74) {$\mu_2$};
    \node[font=\scriptsize,anchor=north] at (0.95,0) {$|\theta|=\frac{|\mu_1-\mu_2|}{|\mu_1-\bar\mu_2|}$};
  \end{scope}
  \node[font=\scriptsize,csig!80!black,anchor=west] at (1.15,-2.4) {signal};
  \node[font=\scriptsize,cidl!80!black,anchor=east] at (-1.15,-2.4) {idler};
  \node[font=\scriptsize,cpmp!70!black,anchor=east] at (-0.12,0.4) {pump};
\end{scope}
\end{tikzpicture}
\caption{\label{fig:pole}Schematic of an event and of a resonance in the $(\tau,z)$ plane, pump frame. (a) Fusion event. Each leg is a soliton of a $2\times2$ Zakharov--Shabat block, the block complementary to its wave, with eigenvalue $\zeta=\mu w_m/2$, Eq.~\eqref{eq:eigmap}; at the vertex (shaded) the three waves overlap and the factorization into blocks fails. The three eigenvalues are images of a single pole $\mu$ of the $3\times3$ problem, which also fixes the common vertex and the locked phases: the event is not a scattering of three pulses, but a single pole that governs all of them. (b) Resonance of a signal and an idler pure pulse with split poles $\mu_1\neq\mu_2$: a fusion followed by a decay, joined by a pump pulse of lifetime $\ell$, Eq.~\eqref{eq:resN3}; the outgoing legs are displaced from free propagation (dotted) by $\Delta\tau_{1,2}$, Eq.~\eqref{eq:delays}.}
\end{figure*}

If one component of $\bm n^0_k$ vanishes, $n^0_{k,m}=0$, the vertex \eqref{eq:vertex} is at infinity and the soliton reduces to a single $\mathrm{sech}$ pulse of wave $m$, propagating undistorted at $v_m$: a soliton of the $2\times2$ Zakharov--Shabat problem of the block complementary to $m$, embedded in the three-wave problem. These \emph{pure pulses} are the soliton content of separated pulses of a single wave (Sec.~\ref{sec:inputface}); a $\mathrm{sech}$ pulse of area $\pi$ is exactly a pure pulse. Whether two pure pulses interact depends on which waves they belong to; pure pulses of the same wave travel at the same velocity and never meet. Figure~\ref{fig:pole} summarizes the two situations treated below: an event, in which a single pole governs the three blocks, and the resonance of two pure pulses whose poles are split.

\emph{Instability.} A pure pump pulse, $\bm n^0=(n^0_1,n^0_2,0)$, is the limit $n^0_3\to0$ of the decay event of the same pole, whose vertex recedes as $\Gamma z_0=-\ln|n^0_3|^2+{\rm const}$ by Eq.~\eqref{eq:vertex}. To first order in $n^0_3$, Eq.~\eqref{eq:1sol} is the pump pulse plus a signal--idler component, in the pump frame and with $y=\tau/T_3$,
\begin{equation}\label{eq:mode}
\begin{pmatrix}\delta u_1\\ \delta\bar u_2\end{pmatrix}\propto\bar n^0_3\,e^{\Gamma z/2}\begin{pmatrix}\sqrt{w_1}\,e^{-ry}\\ \sqrt{w_2}\,e^{(1-r)y}\end{pmatrix}\mathrm{sech}\,y,
\end{equation}
an exact solution of the equations linearized about the pure pulse: a growing mode, with rate $\Gamma/2$ in amplitude and $\Gamma$ in photon number, which is the tail of the event. The instability of a pulse of the intermediate-velocity wave is known \cite{KRB79,Lombardo11}, and for an undepleted $\mathrm{sech}$ pump the growing modes are known in closed form \cite{Gorbach23}: Eq.~\eqref{eq:mode} is their fundamental mode for a pump of area $\pi$, and the pump area $\pi/2$ below which they cease to be localized is the area threshold of Sec.~\ref{sec:inputface}. Eq.~\eqref{eq:mode} adds that its growth ends in a conversion event, so that the depth of the decay follows from the conversion curve. If at depth $z$ the pulse carries along this mode a fraction $\varepsilon\ll1$ of its photons, by $1-J_r(K)\simeq c_+K$ it decays at
\begin{equation}\label{eq:seedlaw}
z_0-z=\frac{1}{\Gamma}\ln\frac{c_+}{\varepsilon}.
\end{equation}
A seed of another shape acts through its projection on the mode, i.e., on the adjoint mode, Eq.~\eqref{eq:mode} with $y\to-y$. Every pump pulse formed in the crystal is such a pulse, and whether and where it decays is decided by $\varepsilon$: the pump leg of an exact fusion event carries no component along the mode, and the unconverted signal and idler it carries decay as $e^{-\Gamma(z-z_0)}$ [Fig.~\ref{fig:faces}(c)].

\begin{figure*}[t]
\includegraphics[width=\textwidth]{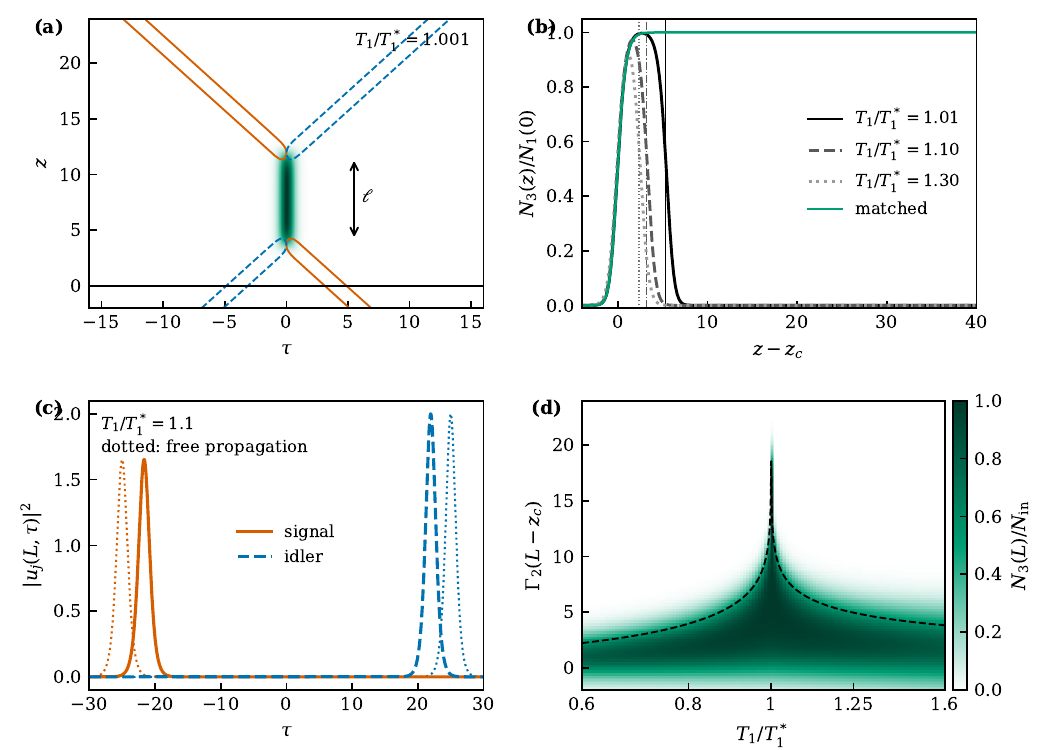}
\caption{\label{fig:sfg}Resonance and fusion of a signal and an idler pure pulse, $\mathrm{sech}$ pulses of area $\pi$; all panels from the degenerate two-pole solution, Eq.~\eqref{eq:resclosed}, with $w_1=w_2=1$ ($r=1/2$, $W=2$), $g=1$ and idler pole $\eta_2=1$ ($\Gamma_2=2$). (a) Solution for $T_1/T_1^*=1.001$, $r=1/2$: pump flux (color) and signal (orange) and idler (blue, dashed) contours. The pump is formed at the collision and decays back after the lifetime $\ell$, Eq.~\eqref{eq:resN3}. (b) Pump photon number in units of the input signal photon number, $N_3(z)/N_1(0)$, for three duration ratios (vertical lines: $\ell$); green: matched durations, single fusion event, Eq.~\eqref{eq:J}. (c) Signal and idler on the output face for $T_1/T_1^*=1.1$, compared with free propagation of the input pulses (dotted): the signal is delayed and the idler advanced by Eq.~\eqref{eq:delays}. (d) Fraction of the input converted to pump on the output face, $N_{\rm in}=\min(N_1,N_2)$, as a function of the duration ratio and of the distance of the face from the collision; $\Gamma_2=2\eta_2w_1w_2$ is the rate of the idler pole, held fixed; dashed: $L-z_c=\ell$, Eq.~\eqref{eq:resN3}.}
\end{figure*}

\subsection{Resonance of signal and idler}\label{sec:resonance}

Let pole 1 be a pure signal pulse ($n^0_{1,1}=0$) and pole 2 a pure idler pulse ($n^0_{2,2}=0$). When they collide, the pump is generated; by Eq.~\eqref{eq:lemma} the dressed vectors remain degenerate in all sectors, so asymptotically no pump survives, and the signal and idler legs are shifted by
\begin{equation}\label{eq:delays}
\Delta\tau_1=\frac{\ln|\theta|^{-1}}{\eta_1w_1},\qquad \Delta\tau_2=-\frac{\ln|\theta|^{-1}}{\eta_2w_2}.
\end{equation}
This is the near-resonant collision of Ref.~\cite{KRB79} (Sec.~IV\,B there), where the two-soliton solution of Ref.~\cite{Kaup76} was used to show that the generated pump decays back, leaving signal and idler delayed, with a delay logarithmic in the mismatch [Eqs.~(4.30)--(4.32) there]. In the present parametrization their two-soliton solution, Eqs.~(4.25)--(4.28) there, takes a simple closed form. With $\bm n_1=(0,a,b)$, $\bm n_2=(c,0,e)$, $\mu_k=i\eta_k$ and $S=\mathbb I$, the Lagrange identity gives $D_1D_2-|p|^2=|a|^2|c|^2+|a|^2|e|^2+|b|^2|c|^2$, so that
\begin{equation}\label{eq:resclosed}
\begin{gathered}
\mathcal D=|ac|^2+|ae|^2+|bc|^2+|\theta|^2|be|^2,\\
\mathcal N_{21}=-\frac{2\eta_1\eta_2}{\eta_1+\eta_2}\,a\bar b\bar ce,\\
\mathcal N_{23}=\eta_1a\bar b\big(|c|^2+\vartheta|e|^2\big),\\
\mathcal N_{31}=\eta_2e\bar c\big(|a|^2-\vartheta|b|^2\big),
\end{gathered}
\end{equation}
with $\vartheta=(\eta_1-\eta_2)/(\eta_1+\eta_2)$, $|\theta|=|\vartheta|$, inserted in Eq.~\eqref{eq:2sol}. The pump, $\propto|abce|/\mathcal D$, is appreciable only where the four terms of $\mathcal D$ are comparable, and the term $|\theta|^2|be|^2$, the only one sensitive to the mismatch, is what eventually terminates it.

The signal, which is faster than the pump, is delayed ($\Delta\tau_1>0$), and the idler, which is slower, is advanced ($\Delta\tau_2<0$): both are shifted toward the group delay of the pump. These time shifts are $\Delta\tau_1=(v_3-v_1)\ell_1$ and $\Delta\tau_2=(v_3-v_2)\ell_2$ with
\begin{equation}\label{eq:lifetime}
\ell_k=\frac{\ln|\theta|^{-1}}{\eta_kw_1w_2}=\frac{2}{\Gamma_k}\ln\frac{1}{|\theta|},\qquad \Gamma_k=2|\eta_k|w_1w_2,
\end{equation}
i.e., signal and idler emerge as if they had travelled at the pump velocity over the distances $\ell_1$ and $\ell_2$. At the collision the two pulses fuse into a pump pulse that propagates over a distance $\ell=r\ell_1+(1-r)\ell_2$, Eq.~\eqref{eq:resN3} below, and then decays back into signal and idler (Fig.~\ref{fig:sfg}). The pump pulse is a \emph{resonance} of the signal--idler pair, and $\ell$ is its lifetime, measured in depth.

\emph{Decay of the resonance.} The pump pulse formed at the collision is a pure pump pulse that carries a component along its growing mode, Eq.~\eqref{eq:mode}, set by the mismatch. Its photon number follows from Eq.~\eqref{eq:resclosed} as the conversion curve follows from Eq.~\eqref{eq:1sol}. With $X=|a/b|^2$ and $Y=|c/e|^2$, $\mathcal D=|be|^2(XY+X+Y+|\theta|^2)$ and $|\mathcal N_{21}|^2/\mathcal D^2\propto XY/(XY+X+Y+|\theta|^2)^2$. Since $X\propto e^{2\eta_1w_1(x+w_2t)}$ and $Y\propto e^{-2\eta_2w_2(x-w_1t)}$, at fixed $t$ the variable $y=Y/X$ is an exponential in $x$ alone, while $Y$ and $|\theta|^2/X$ are powers of $y$ times exponentials in $t$. The integral of $|u_3|^2$ over $x$ is then, exactly,
\begin{gather}
\frac{N_3(z)}{\mathcal N_\theta}=\int_0^\infty\frac{dy}{\big(1+y+K_{\rm f}\,y^{1-r_\theta}+K_{\rm d}\,y^{r_\theta}\big)^2},\label{eq:resN3}\\
K_{\rm f}=e^{-\Gamma_\theta(z-z_{\rm f})},\qquad K_{\rm d}=e^{\Gamma_\theta(z-z_{\rm d})},\nonumber\\
z_{\rm d}-z_{\rm f}=\ell=\frac{2}{\Gamma_\theta}\ln\frac{1}{|\theta|},\nonumber\\
\Gamma_\theta=\frac{2W}{T_1+T_2},\qquad r_\theta=\frac{T_1}{T_1+T_2},\nonumber\\
\mathcal N_\theta=(1-|\theta|^2)\,\frac{W\Gamma_\theta}{g^2},\nonumber
\end{gather}
with $T_k=1/(\eta_kw_k)$ the durations of the two pure pulses and $z_{\rm f,d}$ fixed by the moduli of $\bm n^0_{1,2}$. The same holds for complex poles, with $|\theta|$ of Eq.~\eqref{eq:theta}: $\mathcal D$ keeps the form \eqref{eq:resclosed} and $|\mathcal N_{21}|=2\eta_1\eta_2|abce|/|\mu_1-\bar\mu_2|$. At matching $r_\theta=r$, $\Gamma_\theta=\Gamma$ and $\mathcal N_\theta=\mathcal N$; in general $1/\Gamma_\theta=r/\Gamma_1+(1-r)/\Gamma_2$, so that $\ell=r\ell_1+(1-r)\ell_2$ with $\ell_k$ of Eq.~\eqref{eq:lifetime}. Equation~\eqref{eq:resN3} reduces to the conversion curve \eqref{eq:J} of a fusion at $z_{\rm f}$ where $K_{\rm d}\ll1$ and to that of a decay at $z_{\rm d}$ where $K_{\rm f}\ll1$. The resonance is a fusion event followed by a decay event, joined by a pump pulse: the two vertices of Fig.~\ref{fig:sfg}(a). At the fusion vertex $K_{\rm d}=|\theta|^2$, i.e., the mismatch is a seed of weight $\varepsilon=c_+|\theta|^2$ (with $r\to r_\theta$ in $c_+$), and by Eq.~\eqref{eq:seedlaw}, with $\Gamma\to\Gamma_\theta$, $\ell$ is the depth at which a pump pulse with this seed decays. The time shifts \eqref{eq:delays} follow: between the two vertices the photons travel at the pump velocity, and $\Delta\tau_1-\Delta\tau_2=w_2\ell_1+w_1\ell_2=W\ell$. The pump fused from unmatched pulses thus does not leak back into signal and idler: it stays intact while the seed grows, and decays at once, in a time-reversed copy of the fusion. The lifetime is Eq.~(4.32) of Ref.~\cite{KRB79} up to a factor 2. In the notation of that reference, by Eq.~(4.30) there the outgoing legs cross with a delay $\tau/2$ with respect to the incoming ones, so that $\tau=2\ell$. On the line $z_1\eta_1+z_3\eta_3=0$, which near matching follows the pump pulse, the denominator (4.28) there reads $D=2+e^{2d}+|\theta|^2e^{-2d}$ with $d=z_3\eta_3-z_1\eta_1$, close to its minimum for $\ln|\theta|\lesssim d\lesssim0$. The pump thus forms within a conversion length, stays constant over $\ell=\tau/2$ and decays within a conversion length, as stated by Eq.~\eqref{eq:resN3}; Ref.~\cite{KRB79} describes instead a growth over $\tau/2$ followed by a decay over $\tau/2$.

A pure pulse can be represented by the pole $\mu$ or by $\bar\mu$. For a vector with a vanishing component the rank-two zero at $\bar\mu$ of Sec.~\ref{sec:onesol} splits into a diagonal factor, which does not contribute to $Q$, and a rank-one zero within the $2\times2$ block: unlike for an event, the sign of $\eta$ carries no information. Changing the sign of $\eta_2$ thus describes the same idler pulse, and Eqs.~\eqref{eq:delays} and \eqref{eq:lifetime} are unchanged. Formally, $|\theta|\to1/|\theta|$, but the sectors of pole 2 crossed by the signal leg are exchanged as well, the dominant component on each side of the idler leg being reversed: the shift is then acquired on the output side instead of the input side, with the opposite sign, and the net displacement is the same.

\emph{Matching.} A permanent conversion event requires the signal and idler poles to coincide, which is the resonant pairing condition of Ref.~\cite{KRB79} [Eq.~(4.22) there]. For two $\mathrm{sech}$ pulses of area $\pi$ and durations $T_{1,2}$, Eq.~\eqref{eq:etasech} gives $\eta_1=1/(w_1T_1)$, $\eta_2=1/(w_2T_2)$, and the condition is
\begin{equation}\label{eq:matching}
w_1T_1=w_2T_2,\qquad\text{i.e.}\qquad \frac{T_1}{T_2}=\frac{|\beta_{1,3}-\beta_{1,1}|}{|\beta_{1,2}-\beta_{1,3}|},
\end{equation}
in which case the pair is exactly the fusion soliton of Fig.~\ref{fig:events}; for $r=1/2$ the condition is simply $T_1=T_2$. For arbitrary areas the leading poles coincide when
\begin{equation}\label{eq:matchingh}
\frac{2h_1-1}{w_1T_1}=\frac{2h_2-1}{w_2T_2},
\end{equation}
and all the poles coincide pairwise only if, in addition, $h_1=h_2$: the matching constrains the ratio of the pulse energies as well as that of the durations, and Eq.~\eqref{eq:matchingh}, in closed form, is consistent with the optimal intensity ratio found in Refs.~\cite{IbragimovStruthers96,IbragimovStruthers97}. With equal integer areas, $h_1=h_2=N$, and matched durations both pulses are reflectionless and the pair is an exact $N$-soliton made of $N$ fusion events, each with its own depth and pump duration. Away from matching, with $T_1^*=w_2T_2/w_1$,
\begin{equation}
|\theta|=\frac{|\eta_1-\eta_2|}{\eta_1+\eta_2}=\frac{|T_1/T_1^*-1|}{T_1/T_1^*+1},
\end{equation}
and the generated pump is the resonance described above. This is the decay of the sum-frequency pulse described in Ref.~\cite{Baronio10}, and it is consistent with the time shifts of the fundamental pulses reported in Ref.~\cite{IbragimovStruthers97}. Figure~\ref{fig:sfg} shows the degenerate two-pole solution \eqref{eq:resclosed} for such pulses. With $z_c$ the depth of the collision, Eq.~\eqref{eq:z0fusion}, the pump is found on the output face only if the face lies after the fusion is complete and before the resonance decays,
\begin{equation}\label{eq:window}
\ell_{\rm out}\lesssim L-z_c\lesssim\ell=\frac{T_1+T_2}{W}\ln\frac{T_1/T_1^*+1}{|T_1/T_1^*-1|},
\end{equation}
so that the tolerance on the duration ratio shrinks exponentially with the distance of the face from the collision, $|T_1/T_1^*-1|\lesssim2e^{-\Gamma(L-z_c)/2}$ [Fig.~\ref{fig:sfg}(d)]; outside this window the only trace of the collision is the pair of time shifts \eqref{eq:delays}.

In type-II second-harmonic generation at $r=1/2$ a single pulse polarized at about $45^\circ$ to the crystal axes provides a signal and an idler with equal durations and, for a suitable polarization angle, equal areas, so that Eq.~\eqref{eq:matchingh} is satisfied automatically. The two components enter the crystal superposed, i.e., at their collision point, so that they must be pre-delayed, as in the pulse compressors of Refs.~\cite{WangDragila90,Umbrasas95,Dubietis98}: the faster component enters later by $\tau_1^{\rm in}-\tau_2^{\rm in}=Wz_c$, Eq.~\eqref{eq:z0fusion}, with $z_c\ge\ell_{\rm in}$ for the input face to carry the legs of a fusion event.

\emph{Back-conversion.} A matched fusion event is permanent only as long as its two poles coincide, i.e., as long as the fused pump carries nothing along its growing mode. Anything in the input that splits the poles is a seed $c_+|\theta|^2$, and the fused pump decays back after $\ell=(2/\Gamma)\ln|\theta|^{-1}$, Eq.~\eqref{eq:seedlaw}: it survives over many conversion lengths $\Gamma^{-1}$ for small splittings, but not indefinitely. Noise is expected to split the poles \cite{KRB79}, but it also adds directly a component along the growing mode, through $n_3$; we do not treat its effect quantitatively. Whether the poles are split is decided by the block data of the input face. A signal pulse affects only the $(2,3)$ block and an idler pulse only the $(3,1)$ block, so that for one of each $a_1(\lambda)=\mathcal T^{\rm idler}_{11}(\lambda)$ and $a_2(\lambda)=\mathcal T^{\rm signal}_{22}(\lambda)$, whatever their order \cite{KRB79} (see also Appendix~A of Ref.~\cite{Ibragimov99}). The idler pole is a zero $\lambda_2$ of $a_1$ in the lower half plane, the signal pole a zero $\lambda_1$ of $a_2$ in the upper half plane, matching means $\lambda_1=\bar\lambda_2$, and $\theta=(\lambda_1-\bar\lambda_2)/(\lambda_1-\lambda_2)$. A second, weak signal pulse belongs to the same block, $a_2=(\mathcal T^{\rm seed}\mathcal T^{\rm signal})_{22}$, and moves $\lambda_1$; radiation contained in the input pulses is part of the block data as well. This differs from the plane-wave theory, in which the energy flows periodically between pump and subharmonic waves \cite{Armstrong62}: with walk-off the exchange consists of single conversion events, and a return of the energy requires a new event, i.e., split poles.

\section{Scales and proposed measurements}\label{sec:scales}

\subsection{Scales}

As an example we take type-II interaction in periodically poled KTiOPO$_4$ (PPKTP) at the extended-phase-matching point of Ref.~\cite{KoenigWong04}: signal and idler at $1584$~nm polarized along $y$ and $z$, pump at $792$~nm polarized along $y$. With the Sellmeier equations of Refs.~\cite{KoenigWong04,Fradkin99} the group indices are $1.7638$, $1.8522$ and $1.8078$, so that $W=294.9$~fs/mm, $w_1=148.1$~fs/mm, $w_2=146.8$~fs/mm and $r=0.498$: the pump group velocity is intermediate and essentially the mean of the other two. The group-velocity dispersion is $31$, $51$ and $204$~fs$^2$/mm. In this configuration femtosecond pump pulses at $792$~nm have been down-converted in a $3$-cm crystal with $43\%$ internal efficiency \cite{Pu07}.

With $d_{\rm eff}=2.4$~pm/V and an output pump duration $T_3=300$~fs (sech parameter; full width at half maximum $1.76\,T_3$), the fusion soliton has $T_{1,2}\simeq600$~fs, $\Gamma^{-1}=2.0$~mm, peak intensities $0.49$~GW/cm$^2$ (signal and idler) and $2.0$~GW/cm$^2$ (pump) by Eq.~\eqref{eq:Ipk}, and fluences $0.59$~mJ/cm$^2$ (signal, idler) and $1.2$~mJ/cm$^2$ (pump), i.e., $0.09$~nJ (signal, idler) and $0.19$~nJ (pump) in a waveguide of $16~\mu{\rm m}^2$ effective area. The minimum crystal length for the asymptotic description of both faces, Eq.~\eqref{eq:Lmin}, is $L_{\min}=15.9$~mm for $\delta=10^{-2}$ and $23.4$~mm for $\delta=10^{-3}$. All lengths scale as $T_3$ and the intensities as $T_3^{-2}$: for $T_3=1$~ps, $\Gamma^{-1}=6.8$~mm and $L_{\min}=53$~mm, while for $T_3=200$~fs, $L_{\min}=10.6$~mm. The relevant crystal lengths are thus $1$--$3$~cm for sub-picosecond pulses, within the range of commercially available periodically poled crystals. Table~\ref{tab:observables} collects the quantities accessed by the measurements below, with their values for this example.

For dispersion the relevant parameter is its strength over one conversion length, $\beta_{2,j}\Gamma^{-1}/T_j^2$. With $T_3=300$~fs it is $1.8\times10^{-4}$ for the signal, $2.9\times10^{-4}$ for the idler and $4.6\times10^{-3}$ for the pump, so that dispersion is a weak perturbation of a conversion event. We conjecture that, like other perturbations, it splits coinciding poles by an amount at most of this order; we have not verified this by solving the dispersive equations. If so, since splittings act only logarithmically, $|\theta|\sim10^{-3}$ still leaves a resonance lifetime of about $14\,\Gamma^{-1}$, well beyond $\ell_{\rm out}$, and dispersion bounds the lifetime at exact matching while leaving the quantities of Table~\ref{tab:observables} unchanged for mismatches above this level. The Kerr effect adds to the pump the phase $2\pi n_2I_3/\lambda_3\Gamma=0.03$--$0.08$~rad per conversion length, for a nonlinear index $n_2$ between $0.9$ and $2.4\times10^{-15}$~cm$^2$/W \cite{Li01,DeSalvo96}, and eight times less to signal and idler through self-phase modulation; cross-phase modulation of signal and idler by the pump is of the same order as the self-phase modulation of the pump where the pulses overlap, and is not included. It is negligible on the input pulses of a fusion and stays below $0.5$~rad for a pump that lives less than six conversion lengths; over the ten conversion lengths of a seeded decay it reaches $0.3$--$0.8$~rad, which, treated as a perturbation of the growth rate, lowers it by up to $9\%$. These are order-of-magnitude estimates; a solution of the full equations with dispersion and Kerr terms is needed to quantify both effects.

\begin{table}[b]
\caption{\label{tab:observables}Quantities accessed by the measurements of Secs.~\ref{sec:curve}--\ref{sec:seed}, their values for PPKTP at extended phase matching with $T_3=300$~fs, and the parameter scanned on the output face.}
\begin{ruledtabular}
\begin{tabular}{lll}
quantity & value & scan\\
\hline
$\Gamma^{-1}$, Eq.~\eqref{eq:Gamma} & $2.0$~mm & delay\\
$J_r$, Eq.~\eqref{eq:J} & $J_r(1)=0.53$ & delay\\
$\ell$, Eq.~\eqref{eq:resN3} & $13$~mm\footnotemark[1] & delay\\
$\Delta\tau_{1,2}$, Eq.~\eqref{eq:delays} & $2.0$, $-1.8$~ps\footnotemark[1] & mismatch\\
event shift, Eq.~\eqref{eq:pumpshift} & $0.8$~ps\footnotemark[2] & replica delay\\
$\delta z_0$, Eq.~\eqref{eq:seedlaw} & $4.7$~mm per $10$~dB & seed energy\\
$\delta\tau_1/\delta\tau_2$, Eq.~\eqref{eq:corr} & $-0.99$ & seed energy\\
\end{tabular}
\end{ruledtabular}
\footnotetext[1]{Duration mismatch $T_1/T_1^*=1.1$.}
\footnotetext[2]{Two fusion events with $T_3=300$ and $200$~fs.}
\end{table}

\subsection{Setup}\label{sec:setup}

In the propagation reading every quantity that depends on the depth can be read on the output face of one crystal of fixed length $L$. By Eq.~\eqref{eq:z0fusion} the delay $\Delta=\tau_1^{\rm in}-\tau_2^{\rm in}$ between the input pulses places the vertex at $z_0=\Delta/W$: in PPKTP a step of $10$~fs moves it by $34~\mu$m, and $\Delta=WL=5.9$~ps takes it across a crystal of $20$~mm. A delay scan replaces a scan of the crystal length; for a seeded decay the same role is played by the energy of the seed (Sec.~\ref{sec:seed}). Each measurement below tests a closed-form result of Sec.~\ref{sec:solitons} that concerns the region where the waves overlap and is not contained in the asymptotic description of the faces.

The source is a train of transform-limited $\mathrm{sech}$ pulses at $1584$~nm, as delivered by a soliton mode-locked fiber laser or a synchronously pumped parametric oscillator, with $T_{1,2}\simeq600$~fs (full width $1.06$~ps). A half-wave plate sets the polarization at about $45^\circ$ to the $y$ and $z$ axes of the crystal, and with it the ratio of the areas $h_1/h_2$ of the two components, which are then delayed by $\Delta$ in a polarization interferometer (two polarizing beam splitters and a translation stage). A band-pass filter or a pulse shaper \cite{Weiner00} in one arm changes the duration ratio, and a second translation stage in the same arm provides the replica of Sec.~\ref{sec:shiftmeas}. For the seeded decay of Sec.~\ref{sec:seed} the pump at $792$~nm is the second harmonic of the same source, and the seed an attenuated replica of the $1584$-nm pulse; a Ti:sapphire amplifier at $792$~nm, as in Ref.~\cite{Pu07}, requires a seed from a parametric source. The crystal is a channel waveguide, or a bulk crystal in which the detected region of the beam has uniform intensity, since each transverse point has its own areas (Sec.~\ref{sec:optics}); the energies are those given above. On the output face a dichroic mirror separates the pump, whose energy is read by a photodiode, and a polarizing beam splitter separates signal and idler. The three temporal profiles and arrival times are obtained by sum-frequency cross-correlation with a reference pulse split off before the interferometer. For type-II second-harmonic generation with a single input pulse (Sec.~\ref{sec:resonance}) the pre-delay required for the input face to carry the legs is $Wz_c\gtrsim W\ell_{\rm in}\simeq7T_3\approx2$~ps, and a birefringent plate can replace the interferometer at fixed delay.

\subsection{Conversion curve}\label{sec:curve}

Pulse compression by pre-delayed fundamental pulses has been observed \cite{WangDragila90,Umbrasas95,Dubietis98}, and in type-II second-harmonic generation in KDP the conversion efficiency and the duration of the second harmonic have been measured as functions of the delay, with picosecond pulses of fixed shape and conversion up to $50\%$ \cite{GaneevUsmanov97}. With matched durations and areas $\pi$ the delay dependence is a universal law, not measured so far. The input is then the fusion soliton for $\Delta\ge W\ell_{\rm in}$, and by Eq.~\eqref{eq:J} the pump photon number on the output face is
\begin{equation}\label{eq:delayscan}
\frac{N_3(L)}{\mathcal N}=J_r\big(e^{-\Gamma(L-\Delta/W)}\big),
\end{equation}
also for $\Delta>WL$, when the face cuts the event before its vertex. The pump energy as a function of the delay is thus the conversion curve, with no free parameter once $\Gamma$ is fixed by the pump duration. It passes through $J_r(1)\simeq0.53$ at $\Delta=WL$ and is not symmetric about this point: the unconverted fraction decays as $e^{-\Gamma(L-\Delta/W)}$ on one side, the converted one as $e^{-\Gamma(\Delta/W-L)/\max(r,1-r)}$ on the other [Fig.~\ref{fig:faces}(c)], with $e$-folding delays $W/\Gamma=600$~fs and about $300$~fs in PPKTP.

\subsection{Resonance: lifetime and time shifts}\label{sec:resmeas}

The decay of the generated sum-frequency field has been observed with non-collinear beams in KTiOPO$_4$, as a function of the input intensity \cite{Baronio09}, and the time shifts of the fundamental pulses are known theoretically \cite{KRB79,IbragimovStruthers97}; the dependence of the lifetime on the mismatch, Eq.~\eqref{eq:resN3}, and the time shifts have not been measured. A filter in one arm gives pure pulses with $|\theta|=|T_1/T_1^*-1|/(T_1/T_1^*+1)$. More simply, a rotation of the half-wave plate by $\alpha$ from the matched angle changes the areas to $h_{1,2}\simeq1\mp2\alpha$ and by Eq.~\eqref{eq:etasech} splits the poles by $|\theta|\simeq4|\alpha|$, with radiation of relative energy $O(\alpha^2)$. The pump is then found on the output face only inside the window \eqref{eq:window}: as a function of the delay its energy is a plateau of width
\begin{equation}
W\ell=(T_1+T_2)\ln\frac{1}{|\theta|}=\Delta\tau_1-\Delta\tau_2,
\end{equation}
i.e., $2.0$, $2.8$ and $3.8$~ps ($\ell=6.7$, $9.5$ and $13.0$~mm) for $T_1/T_1^*=2$, $1.3$ and $1.1$ at fixed $T_2$, which tests the logarithmic law within a crystal of $20$--$30$~mm; the full dependence on the duration ratio and on the delay is the map of Fig.~\ref{fig:sfg}(d). For delays beyond the plateau the pump is absent and signal and idler leave the crystal with their input shapes, delayed and advanced by $T_{1,2}\ln|\theta|^{-1}$, Eq.~\eqref{eq:delays}: $2.0$ and $1.8$~ps for $T_1/T_1^*=1.1$, three pulse durations. Measured together with the plateau width, the time shifts test the same logarithm, and their sum equals the width with no free parameter. At exact matching the lifetime is bounded by the residual asymmetry of the group velocities and, presumably, by dispersion and by the fluctuations of the input: with equal durations, $(w_1-w_2)/W=4.5\times10^{-3}$ gives $\Gamma\ell\simeq2\ln(1/4.5\times10^{-3})\simeq11$, which the plate angle compensates.

\subsection{Shifts between events}\label{sec:shiftmeas}

Two events are prepared by sending through the interferometer two replicas of the input pulse with different durations and areas $\pi$, separated by more than their durations, so that each replica forms a fusion event, $a$ and $b$. Without interaction the two pump pulses would leave the crystal at the weighted means of the input times. By the asymptotics of the two-soliton solution, pulses of the same wave are pushed apart by $T_i\ln|\theta_{ab}|^{-1}$ each on the input and on the output face; since $(1-r)T_1=rT_2=T_3$, the separation of the two pump pulses on the output face is smaller than its kinematic value by
\begin{equation}\label{eq:pumpshift}
\big(T_{3a}+T_{3b}\big)\ln\frac{1}{|\theta_{ab}|},
\end{equation}
for events in sequence, i.e., when signal, idler and pump of $b$ all follow those of $a$. For $T_{3a}=300$~fs and $T_{3b}=200$~fs, $|\theta_{ab}|=0.2$ and the pump pulses are $0.8$~ps closer than the input times imply. The displacement is acquired where the legs cross (Fig.~\ref{fig:two}): it is absent if the crossings lie beyond the output face, which the delay between the replicas controls, and it vanishes when the carrier of one replica is detuned by more than its bandwidth, $\ln|\theta_{ab}|^{-1}\simeq2\eta_a\eta_b/(\xi_a-\xi_b)^2$. Ensembles of non-interacting triads have been observed with beams \cite{Baronio11}; the mutual displacement of interacting events has not, to our knowledge, been measured.

\subsection{Seeded decay and correlated legs}\label{sec:seed}

That the depth of a seeded decay is logarithmic in the seed is known, with its coefficient for a noise seed \cite{KRB79} (Sec.~IV\,A and Appendix~F there); Eq.~\eqref{eq:seedlaw} gives the coefficient, $\Gamma^{-1}$, for a seed along the growing mode and relates the offset to the weight $\varepsilon$ of the seed along the growing mode, which for a seed of given shape requires its projection on the adjoint mode. The measurement below tests the coefficient. The trapping of signal and idler by the gain of a pump pulse with symmetric walk-off, described by modes of the type \eqref{eq:mode} \cite{Gorbach23}, has been observed in lithium niobate nanowaveguides, where it breaks up at saturation \cite{Hamrouni24}, and as a gain-induced reduction of the delay between signal and idler in PPKTP \cite{Thekkadath24}; the depth of the decay as a function of the seed and the correlation of the arrival times have not been measured. A pump pulse at $792$~nm with $T_3=300$~fs and area $\pi$ in its block enters the crystal together with a weak signal seed of energy $E_s$, and the depth of the decay moves as $\delta z_0=-\Gamma^{-1}\delta\ln E_s$: an attenuation of $10$~dB moves it $4.7$~mm deeper. By Eq.~\eqref{eq:corr} signal and idler then reach the output face displaced by $+0.69$ and $-0.69$~ps in the pump frame, in the fixed ratio $-r/(1-r)$, whatever the seed. The measurement is the pair of arrival times as a function of the seed energy; with pulse-to-pulse fluctuations of the seed the two arrival times fluctuate along the same line.

\section{Conclusions}\label{sec:conclusions}

We have shown that the solitons of the three-wave interaction with a pump of intermediate group velocity are conversion events. The usual picture of a soliton, an asymptotic profile held together by the nonlinearity against dispersive spreading and relaxation, does not apply to them: away from the vertex each leg propagates by linear transport, and its shape, timing and phases are fixed where the three waves overlap and the $3\times3$ problem does not factorize. The TWI solitons of the factorized description are the legs of such events.

A pump pulse is unstable, its growing mode is the tail of a decay event, and it decays at a depth logarithmic in the weight of the seed along this mode, Eq.~\eqref{eq:seedlaw}. Unmatched signal and idler pulses fuse into a pump pulse seeded by the splitting of their poles: the resonance they form is a fusion followed by a decay, with the pump photon number \eqref{eq:resN3}: the pump forms and decays within a conversion length each and survives unchanged in between, over $\tau/2$, with $\tau$ the delay of Ref.~\cite{KRB79}, and back-conversion of a matched fusion is the same decay, seeded by whatever splits the poles. For a single event the conversion curve gives the distance from the vertex beyond which the faces carry the legs, and the $N$-soliton is a network of events, displaced along the characteristics of definite waves by pairwise, additive shifts that arise from transient conversion where legs cross.

These quantities belong to the region where the waves overlap and are not contained in the asymptotic description, but the propagation reading makes them accessible from outside. The delay between the input pulses sets the depth of the vertex, and the conversion curve, the lifetime of the resonance, the time shifts, the displacement of one event by another and the correlation of the arrival times of signal and idler, Eq.~\eqref{eq:corr}, are read on the output face of a crystal of a few centimeters (Sec.~\ref{sec:scales}). A pump of intermediate group velocity is required, as in type-II interaction in PPKTP at extended phase matching, where femtosecond pump pulses have already been down-converted \cite{Pu07}. The spatial experiment of Ref.~\cite{Baronio09}, with non-collinear beams in KTiOPO$_4$, observed the decay of the generated sum-frequency beam above an intensity threshold, and the trapping of signal and idler by the gain of the pump has been observed in lithium niobate nanowaveguides \cite{Hamrouni24}; the temporal, collinear setup proposed here adds access to the depth of the vertex through the input delay, and with it to the quantities derived above.

\begin{acknowledgments}
We acknowledge support from HORIZON-ERC-2023-ADG HYPERSPIM project Grant No.~101139828.
\end{acknowledgments}

\emph{Data availability.}---The numerical codes and the data that support the findings of this article are available from the authors upon reasonable request.

\appendix
\section{$\mathfrak r$-matrix and conserved charges from the monodromy}\label{app:charges}

\subsection{Classical $\mathfrak r$-matrix}

The Poisson brackets of $U$ are ultralocal and given by a rational $\mathfrak r$-matrix,
\begin{multline}
\{U(x,\lambda)\overset{\otimes}{,}U(y,\mu)\}\\
=\big[\mathfrak r(\lambda-\mu),\,U(x,\lambda)\otimes\mathbb{I}+\mathbb{I}\otimes U(y,\mu)\big]\delta(x-y),
\end{multline}
\begin{gather}
\mathfrak r(\lambda)=\frac{\kappa}{\lambda}\,\Perm,\qquad \Perm=\sum_{a,b=1}^3E_{ab}\otimes E_{ba},
\end{gather}
where $E_{ab}$ are the matrix units. This is equivalent to the statement that the only nonzero brackets among the entries of $Q$ are
\begin{equation}
\{Q_{jk}(x),Q_{kj}(y)\}=-i\kappa\,(v_k-v_j)\,\delta(x-y).
\end{equation}
For the monodromy $\mathcal T_L(\lambda)=\overleftarrow{\exp}\int_{-L}^{L}U(x,\lambda)\,dx$ one obtains
\begin{equation}\label{eq:rTT}
\{\mathcal T_L(\lambda)\overset{\otimes}{,}\mathcal T_L(\mu)\}=\big[\mathfrak r(\lambda-\mu),\,\mathcal T_L(\lambda)\otimes \mathcal T_L(\mu)\big],
\end{equation}
so that $\{\tr \mathcal T_L(\lambda),\tr \mathcal T_L(\mu)\}=0$; $\mathfrak r$ is Yang's $\mathfrak{gl}(3)$ solution of the classical Yang--Baxter equation \cite{FT}.

\subsection{Charges from the monodromy matrix}

The derivation follows the nonlinear Schr\"odinger case \cite{FT}. With periodic conditions, $u_j(x+2L,t)=u_j(x,t)$, the monodromy $\mathcal T_L(\lambda)$ evolves by conjugation, so its whole spectrum is conserved, and by Eq.~\eqref{eq:rTT} the traces of its powers are in involution. A gauge transformation $\mathcal G=\mathbb I+\Xi$, $\Xi$ off diagonal, makes $U$ diagonal, $\mathcal G^{-1}(U\mathcal G-\partial_x\mathcal G)=\diag(d_1,d_2,d_3)$, provided $\Xi$ solves the matrix Riccati equation
\begin{equation}
\partial_x\Xi=-i\lambda[A,\Xi]+Q+(Q\Xi)^{\rm off}-\Xi(Q\Xi)^{\rm diag},
\end{equation}
with $d_j=-i\lambda v_j+(Q\Xi)_{jj}$. The eigenvalues of $\mathcal T_L$ are then $\exp\int_{-L}^{L}d_j\,dx$. Expanding $\Xi=\sum_{n\ge1}\Xi_n\lambda^{-n}$ and $d_j=-i\lambda v_j+\sum_{n\ge1}d_j^{(n)}\lambda^{-n}$ gives $(\Xi_1)_{jk}=Q_{jk}/[i(v_j-v_k)]$, the recursion
\begin{equation}
i[A,\Xi_{n+1}]=\big(Q\Xi_n\big)^{\rm off}-\partial_x\Xi_n-\!\!\sum_{\substack{m+p=n\\ m,p\ge1}}\!\!\Xi_m\big(Q\Xi_p\big)^{\rm diag},
\end{equation}
and $d^{(n)}_j=(Q\Xi_n)_{jj}$, local in the fields. With $I^{(n)}_j=\int d^{(n)}_j\,dx$, $p_k=-i\int dx\,\bar u_k\partial_xu_k$, $\mathcal V=\int dx\,(\bar u_1\bar u_2u_3+u_1u_2\bar u_3)$ and indices mod 3,
\begin{gather}
I^{(1)}=i\kappa\big(N_2+N_3,\ -N_1-N_3,\ N_1-N_2\big),\\
I^{(2)}_j=i\kappa\Big[\frac{p_{j-1}}{w_{j-1}}-\frac{p_{j+1}}{w_{j+1}}+\frac{g\,w_j}{w_1w_2w_3}\,\mathcal V\Big],\\
P=-\frac{i}{\kappa}\sum_jA_{jj}I^{(2)}_j,\qquad H=\frac{i}{\kappa}\sum_jB_{jj}I^{(2)}_j .
\end{gather}
Since $\det \mathcal T_L=e^{-2i\lambda L\sum_jv_j}$, $\sum_jI^{(n)}_j=0$, and since $\{(1),(v_j),(v_kv_l)\}$ is a basis of $\mathbb C^3$ for distinct velocities, each order contains at most two independent charges: the Manley--Rowe charges $\mathcal Q_{1,2}$ at $n=1$, and $P$ and $H$ at $n=2$. The cubic vertex survives in $H$ but not in $P$ because $\sum_jB_{jj}w_j=-w_1w_2w_3$ while $\sum_jA_{jj}w_j=0$. For rapidly decreasing fields the same expansion holds for $\ln\mathcal T_{jj}(\lambda)$, the $\mathfrak{gl}(3)$ analog of $\ln a(\lambda)$.

\bibliography{bibliografia_classical_v2}

\end{document}